%% file: DGSlepton2Lep.tex
\documentclass[11pt,a4paper]{article}

\usepackage{jheppub}

\usepackage{preprintcover}  
\PreprintCoverPaperTitle{%
Search for direct production of charginos, neutralinos and sleptons 
in final states with two leptons and missing transverse momentum 
in $pp$ collisions at $\sqrt{s}$~=~8\,TeV with the ATLAS detector}
\PreprintIdNumber{CERN-PH-EP-2014-037}
\PreprintCoverAbstract{%
Searches for the electroweak production of charginos, neutralinos and sleptons in final 
states characterized by the presence of two leptons (electrons and muons) and missing 
transverse momentum are performed using \datalumi\ of proton-proton collision data 
at $\sqrt{s}=8\TeV$ recorded with the ATLAS experiment at the Large Hadron Collider. 
No significant excess beyond Standard Model expectations is observed. 
Limits are set on the masses of the lightest chargino, next-to-lightest neutralino and 
sleptons for different lightest-neutralino mass hypotheses in simplified models. 
Results are also interpreted in various scenarios of the phenomenological Minimal Supersymmetric Standard Model.}
\PreprintJournalName{JHEP}

\usepackage{atlasphysics}

\usepackage{epstopdf}
\usepackage{array}
\usepackage{hyperref} 

\newcommand{\datalumi}{20.3\,\ifb}
\newcommand{\lumierror}{2.8\%}
\newcommand{\exclMslepMinZero}{90\,GeV}      
\newcommand{\exclMslepMaxZero}{325\,GeV}     
\newcommand{\exclMslepMinHundred}{160\,GeV}  
\newcommand{\exclMslepMaxHundred}{310\,GeV}  
\newcommand{\exclMconeMinZero}{140\,GeV}     
\newcommand{\exclMconeMaxZero}{465\,GeV}     
\newcommand{\exclMntwoMinZero}{180\,GeV}     
\newcommand{\exclMntwoMaxZero}{355\,GeV}     
\newcommand{\exclMntwoMinZeroComb}{100\,GeV} 
\newcommand{\exclMntwoMaxZeroComb}{415\,GeV} 

\newcommand{\alpgen}{{\tt ALPGEN}}
\newcommand{\powheg}{{\tt POWHEG}}
\newcommand{\prospino}{{\tt PROSPINO}}
\newcommand{\herwig}{{\tt HERWIG}}
\newcommand{\pythia}{{\tt PYTHIA}}
\newcommand{\mcatnlo}{{\tt MC@NLO}}
\newcommand{\amcnlo}{\texttt{aMC@NLO}}

\newcommand{\dynnlo}{{\tt DYNNLO}}
\newcommand{\mcfm}{{\tt MCFM}}

\newcommand{\jimmy}{{\tt JIMMY}}
\newcommand{\sherpa}{{\tt SHERPA}}

\newcommand{\acer}{{\tt AcerMC}}
\newcommand{\madgraph}{\texttt{MADGRAPH}}

\newcommand{\mttwo}{\ensuremath{m_\mathrm{T2}}}

\makeatletter
\g@addto@macro\bfseries{\boldmath}
\makeatother

\newcommand{\real}{prompt}

\newcommand{\fake}{non-prompt}
\newcommand{\Fake}{Non-prompt}
\newcommand{\reff}{\ensuremath{\epsilon_\mathrm{p}}}
\newcommand{\feff}{\ensuremath{\epsilon_\mathrm{n}}}

\newcommand{\gaugino}[2]{\ensuremath{\tilde\chi_{#1}^{#2}}}
\newcommand{\neutralino}[1]{\gaugino{#1}{0}}
\newcommand{\chargino}[1]{\gaugino{#1}{\pm}}
\newcommand{\Cone}{\chargino{1}}
\newcommand{\None}{\neutralino{1}}
\newcommand{\Ntwo}{\neutralino{2}}
\newcommand{\ConeCone}{\gaugino{1}{+}\gaugino{1}{-}}
\newcommand{\ConeNtwo}{\Cone\Ntwo}
\newcommand{\mNone}{\ensuremath{m_{\smash{\neutralino{1}}}}}
\newcommand{\mNtwo}{\ensuremath{m_{\smash{\neutralino{2}}}}}
\newcommand{\mCone}{\ensuremath{m_{\smash{\chargino{1}}}}}
\newcommand{\mslepton}{\ensuremath{m_{\smash{\slepton}}}}
\newcommand{\msneutrino}{\ensuremath{m_{\smash{\sneutrino}}}}
\newcommand{\sneutrino}{\ensuremath{\tilde\nu}}
\newcommand{\gravitino}{\ensuremath{\tilde{G}}}
\newcommand{\pTmiss}{\ensuremath{\mathbf{p}_\mathrm{T}^\mathrm{miss}}}
\newcommand{\METrel}{\ensuremath{E_\mathrm{T}^\mathrm{miss,rel}}}
\newcommand{\mll}{\ensuremath{m_{\ell\ell}}}
\newcommand{\pTll}{\ensuremath{p_\mathrm{T,\ell\ell}}}

\newcommand{\selectron}{\ensuremath{\tilde{e}}}
\newcommand{\smuon}{\ensuremath{\tilde{\mu}}}
 
\newcommand{\limsigvis}{\ensuremath{\sigma_\mathrm{vis}^{95}}} 
\newcommand{\CLs}{CL$_\mathrm{s}$}
\newcommand{\SRmttwo}{SR-\mttwo}
\newcommand{\SRmtt}[1]{SR-$m_\mathrm{T2}^{#1}$}
\newcommand{\SRmtta}{\SRmtt{90}}

\newcommand{\SRWW}{SR-$WW$}
\newcommand{\SRWWa}{\SRWW a}
\newcommand{\SRWWb}{\SRWW b}
\newcommand{\SRWWc}{\SRWW c}
\newcommand{\SRZjets}{SR-$Z$jets}
\newcommand{\emu}{\ensuremath{e^\pm\mu^\mp}}
\newcommand{\dRll}{\ensuremath{\Delta R_{\ell\ell}}}
\newcommand{\mjj}{\ensuremath{m_{jj}}}

\newcolumntype{L}{>{$}l<{$}}
\newcolumntype{C}{>{$}c<{$}}
\newcolumntype{R}{>{$}r<{$}}

\newcommand{\spz}{\phantom{0}}

\newcommand{\plot}[3]{\includegraphics[width=#1\textwidth]{#2}}

\begin{document}


\title{Search for direct production of charginos, neutralinos and sleptons 
in final states with two leptons and missing transverse momentum 
in $pp$ collisions at $\sqrt{s}$~=~8\,TeV with the ATLAS detector}

\author{The ATLAS Collaboration}

\abstract{%
Searches for the electroweak production of charginos, neutralinos and sleptons in final 
states characterized by the presence of two leptons (electrons and muons) and missing 
transverse momentum are performed using \datalumi\ of proton-proton collision data 
at $\sqrt{s}=8\TeV$ recorded with the ATLAS experiment at the Large Hadron Collider. 
No significant excess beyond Standard Model expectations is observed. 
Limits are set on the masses of the lightest chargino, next-to-lightest neutralino and 
sleptons for different lightest-neutralino mass hypotheses in simplified models. 
Results are also interpreted in various scenarios of the phenomenological Minimal Supersymmetric Standard Model.
}

\maketitle


\section{Introduction}

Supersymmetry (SUSY)~\cite{Miyazawa:1966,Ramond:1971gb,Golfand:1971iw,Neveu:1971rx,Neveu:1971iv,Gervais:1971ji,Volkov:1973ix,Wess:1973kz,Wess:1974tw} is a spacetime symmetry
that postulates for each Standard Model (SM) particle the existence of a partner particle whose spin differs by one-half unit.
The introduction of these new particles provides a potential solution to the hierarchy problem~\cite{Weinberg:1975gm,Gildener:1976ai,Weinberg:1979bn,Susskind:1978ms}.
If $R$-parity is conserved~\cite{Fayet:1976et,Fayet:1977yc,Farrar:1978xj,Fayet:1979sa,Dimopoulos:1981zb},  
as is assumed in this paper, SUSY particles are always produced in pairs and the lightest supersymmetric particle (LSP)
emerges as a stable dark-matter candidate.

The charginos and neutralinos are mixtures of the bino, winos and
higgsinos that are superpartners of the U(1), SU(2) gauge bosons and
the Higgs bosons, respectively.
Their mass eigenstates are referred to as \chargino{i} $(i=1,2)$ and \neutralino{j} $(j=1,2,3,4)$
in the order of increasing masses.
Even though the gluinos and squarks are produced strongly in $pp$ collisions,
if the masses of the gluinos and squarks are large, the direct production of charginos, neutralinos and sleptons through electroweak interactions
may dominate the production of SUSY particles at the Large Hadron Collider (LHC)\@. 
Such a scenario is possible in the general framework of the phenomenological minimal supersymmetric SM (pMSSM)~\cite{pmssm1,pmssm2,pmssm3}.
Naturalness suggests that third-generation sparticles and some of the charginos and neutralinos should have masses of a few hundred GeV~\cite{Barbieri:1987fn,deCarlos:1993yy}.
Light sleptons are expected in gauge-mediated~\cite{Dine:1981gu,AlvarezGaume:1981wy,Nappi:1982hm,Dine:1993yw, Dine:1994vc,Dine:1995ag} 
and anomaly-mediated~\cite{Randall:1998uk,Giudice:1998xp}
SUSY breaking scenarios. 
Light sleptons could also play a role in the co-annihilation of neutralinos, 
allowing a dark matter relic density consistent with cosmological observations~\cite{Belanger:2004ag,King:2007vh}.

This paper presents searches for electroweak production of charginos, neutralinos and sleptons
using \datalumi\ of proton-proton collision data with a centre-of-mass energy $\sqrt{s}=8\TeV$ collected at the LHC with the ATLAS detector.
The searches  target final states with two oppositely-charged leptons
(electrons or muons) and missing transverse momentum.
Similar searches~\cite{Aad:2012pxa,cmsEWK7}
have been performed using $\sqrt{s}=7\TeV$ data 
by the ATLAS and CMS experiments.
The combined LEP limits on the selectron, smuon and chargino masses are
$m_{\selectron}>99.9\GeV$, $m_{\smuon}>94.6\GeV$ and
$m_{\chargino{1}}>103.5\GeV$~\cite{lepsusy,alephsusy,delphisusy,l3susy,opalsusy}.
The LEP selectron limit assumes gaugino mass unification
and cannot be directly compared with the results presented here.

\section{SUSY scenarios}

Simplified models~\cite{Alwall:2008ag} are considered for optimization of the event selection
and interpretation of the results.
The LSP is the lightest neutralino \None\ in all SUSY scenarios considered,
except in one scenario in which it is the gravitino \gravitino.
All SUSY particles except for the LSP are assumed to decay promptly.
In the electroweak production of \ConeCone\ and \ConeNtwo,
\Cone\ and \Ntwo\ are assumed to be pure wino and mass degenerate, and
only the $s$-channel production diagrams, $q\bar{q}\to (Z/\gamma)^*\to\ConeCone$
and $q\bar{q}'\to W^{\pm*}\to\ConeNtwo$, are considered.
The cross-section for $\gaugino{1}{+}\gaugino{1}{-}$ production is 6\,pb for 
a \chargino{1} mass of 100\,GeV and decreases to 10\,fb at 450\,GeV.
The cross-section for \chargino{1}\neutralino{2} production is 11.5\,pb for
a degenerate \chargino{1}/\neutralino{2} mass of $100\GeV$, and 40\,fb for $400\GeV$.

\begin{figure}\centering
	\raisebox{0.15\textwidth}{(a)}\includegraphics[width=0.35\textwidth]{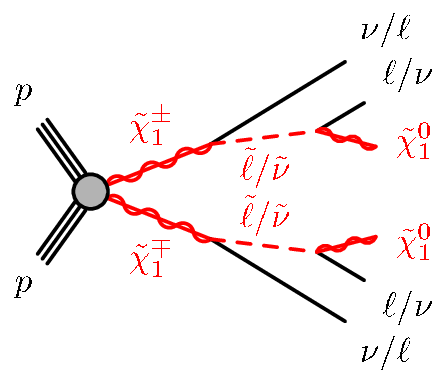}\hfil
	\raisebox{0.15\textwidth}{(b)}\includegraphics[width=0.35\textwidth]{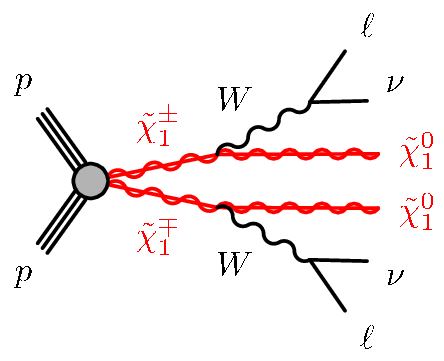}
	\raisebox{0.15\textwidth}{(c)}\includegraphics[width=0.35\textwidth]{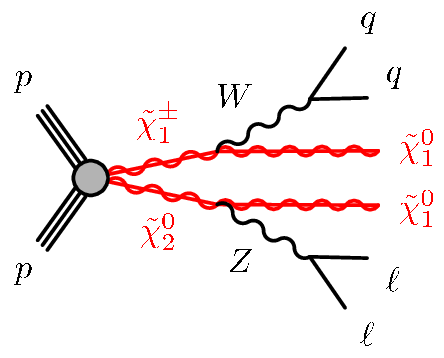}\hfil
	\raisebox{0.15\textwidth}{(d)}\includegraphics[width=0.35\textwidth]{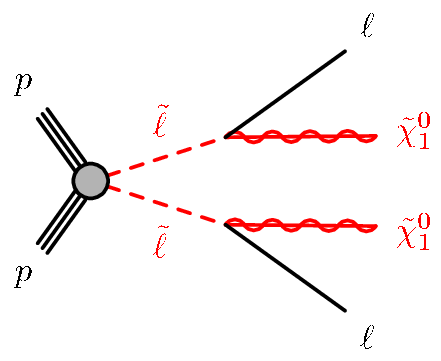}
	\caption{\label{diagrams}
		Electroweak SUSY production processes of the considered simplified models.}
\end{figure}

In the scenario in which 
the masses of the sleptons and sneutrinos lie between the \Cone\ and \None\ masses,
the \Cone\ decays predominantly as 
$\chargino{1} \to (\slepton^\pm\nu$ or $\ell^\pm\sneutrino) \to \ell^\pm\nu\neutralino{1}$.
Figure~\ref{diagrams}(a) shows direct chargino-pair production,
$pp\to\gaugino{1}{+}\gaugino{1}{-}$,
followed by the slepton-mediated decays.
The final-state leptons can be either of the same flavour
(SF = $e^+e^-$ or $\mu^+\mu^-$), or of different flavours (DF = \emu).
In this scenario,
the masses of the three left-handed sleptons and three sneutrinos are assumed to be
degenerate with $\mslepton=\msneutrino=(\mNone+\mCone)/2$.
The \chargino{1} is assumed to decay 
with equal branching ratios (1/6)
into $\slepton^{\pm}\nu$ and $\ell^{\pm}\sneutrino$
for three lepton flavours,
followed by $\slepton^{\pm}\to\ell^{\pm}\neutralino{1}$ or $\sneutrino\to\nu\neutralino{1}$
with a 100\% branching ratio.

In the scenario in which the \Cone\ is the next-to-lightest supersymmetric particle (NLSP),
the \Cone\ decays as $\chargino{1}\to W^\pm\neutralino{1}$.
In direct \ConeCone\ production, if both $W$ bosons decay leptonically
as shown in figure~\ref{diagrams}(b),
the final state  contains two opposite-sign
leptons, either SF or DF, and large missing transverse momentum.

Another scenario is considered in which \Cone\ and \Ntwo\ are mass degenerate and are co-NLSPs.
The direct $\chargino{1}\neutralino{2}$ production is 
followed by the decays $\chargino{1}\to W^{\pm}\neutralino{1}$ and $\neutralino{2}\to Z\neutralino{1}$
with a 100\% branching fraction.
If the $Z$ boson decays leptonically and the $W$ boson decays hadronically,
as shown in figure~\ref{diagrams}(c),
the final state contains two opposite-sign leptons, two hadronic jets, and missing
transverse momentum.
The leptons in this case are SF
and their invariant mass is consistent with the $Z$ boson mass.
The invariant mass of the two jets from the $W$ decay gives an additional 
constraint to characterize this signal.

A scenario in which the slepton is the NLSP is modelled 
according to ref.~\cite{AbdusSalam2011}.
Figure~\ref{diagrams}(d) shows direct slepton-pair production
$pp\to\slepton^+\slepton^-$ followed by
$\slepton^\pm \to \ell^\pm \neutralino{1}$ ($\ell=e$ or $\mu$),
giving rise to a pair of SF leptons and
missing transverse momentum due to the two neutralinos.
The cross-section for direct slepton pair production in this scenario 
decreases from 127\,fb to 0.5\,fb per slepton flavour for left-handed sleptons,
and from 49\,fb to 0.2\,fb for right-handed sleptons,
as the slepton mass increases from 100 to $370\GeV$.

Results are also interpreted in dedicated pMSSM~\cite{Djouadi:1998di} scenarios. 
In the models considered in this paper, the masses of the coloured sparticles, 
of the CP-odd Higgs boson, and of the left-handed sleptons are set to high values to 
allow only the direct production of charginos and neutralinos via $W$/$Z$, and 
their decay via right-handed sleptons, gauge bosons and the lightest Higgs boson.
The lightest Higgs boson mass is set close to $125\GeV$~\cite{Aad:2012tfa,Chatrchyan:2012ufa} 
by tuning the mixing in the top squark sector.
The mass hierarchy, composition and production cross-section of the charginos and neutralinos
are governed by the ratio $\tan\beta$ of the expectation values of the two Higgs doublets, 
the gaugino mass parameters $M_1$ and $M_2$, and the higgsino mass parameter $\mu$.
Two classes of pMSSM scenarios are studied on a $\mu$-$M_2$ grid, distinguished by 
the masses of the right-handed sleptons $\slepton_R$.
If $m_{\smash{\tilde{\ell}_R}}$ lies halfway between $\mNone$ and $\mNtwo$,
\Ntwo\ decays preferentially through $\Ntwo\to\slepton_R\ell\to\None\ell\ell$.
The parameter $\tan\beta$ is set to 6 yielding comparable branching ratios into each slepton generation. 
To probe the sensitivity to different $\ninoone$ compositions, 
three values of $M_1=100$, 140 and $250\GeV$ are considered. 
If, on the other hand, all sleptons are heavy, 
\Cone\ and \Ntwo\ decay via $W$, $Z$ and Higgs bosons.
The remaining parameters are fixed to $\tan\beta=10$ and $M_1=50\GeV$ so that the
relic dark-matter density is below the cosmological bound across the
entire $\mu$-$M_2$ grid.
The lightest Higgs boson has a mass close to $125\GeV$
and decays to both SUSY and SM particles where kinematically allowed. 

In addition,
the gauge-mediated SUSY breaking (GMSB) 
model proposed in ref.~\cite{Meade:6888v2} is considered.
In this simplified model, the LSP is the gravitino $\gravitino$, the NLSP is the chargino with 
$m_{\smash{\chargino{1}}}=110\GeV$, and in addition there are two other light neutralinos 
with masses
$m_{\smash{\neutralino{1}}}=113\GeV$ and $m_{\smash{\neutralino{2}}}=130\GeV$.
All coloured sparticles are assumed to be very heavy.
The $\gaugino{1}{+}\gaugino{1}{-}$ production cross-section is not large ($\sim$1.4\,pb),
but
the same final state is reached via production of $\chargino{1}\neutralino{1}$ ($\sim$2.5\,pb),
$\chargino{1}\neutralino{2}$ ($\sim$1.0\,pb) and
$\neutralino{1}\neutralino{2}$ ($\sim$0.5\,pb).
The $\neutralino{1}$ decays into $\chargino{1}W^{\mp*}$, and the \Ntwo\ decays either
into $\chargino{1}W^{\mp*}$ or $\neutralino{1}Z^*$.
Because of the small mass differences, decay products of the off-shell $W$ and $Z$ bosons
are unlikely to be detected.
As a result, all of the four production channels result in the same experimental signature,
and their production cross-sections can be added together for the purpose of this search.
Each $\chargino{1}$ then decays via  $\chargino{1}\to W^\pm\gravitino$, and leptonic decays of
the two $W$ bosons produce the same final-state as in the other scenarios.

\section{The ATLAS detector}

The ATLAS detector~\cite{Aad:2008zzm} is a multi-purpose particle
physics detector with a forward-backward symmetric cylindrical
geometry and nearly 4$\pi$ coverage in solid angle\footnote{%
	ATLAS uses a right-handed coordinate system with its origin at the nominal 
	interaction point (IP) in the centre of the detector, and the $z$-axis along the 
	beam line. The $x$-axis points from the IP to the centre of the LHC ring, 
	and the $y$-axis points upwards. Cylindrical coordinates $(r,\phi)$ are used 
	in the transverse plane, $\phi$ being the azimuthal angle around the $z$-axis. 
	Observables labelled `transverse' are projected into the $x$--$y$ plane. 
	The pseudorapidity is defined in terms of the polar angle $\theta$ by $\eta=-\ln\tan(\theta/2)$.}.
It contains four superconducting magnet systems, which include a 
thin solenoid surrounding the inner tracking detector (ID),
and barrel and end-cap toroids as part of  
a muon spectrometer (MS). The ID covers the pseudorapidity region  $|\eta| <2.5$ and consists of a 
silicon pixel detector, a silicon microstrip detector, and a transition
radiation tracker. In the pseudorapidity
region $|\eta| < 3.2$, high-granularity liquid-argon (LAr)
electromagnetic (EM) sampling calorimeters are used.  An iron-scintillator tile
calorimeter provides coverage for hadron detection over
$|\eta|<1.7$.  The end-cap and forward regions, spanning
$1.5<|\eta|<4.9$, are instrumented with LAr calorimeters
for both EM and hadronic measurements. The MS surrounds the calorimeters and consists 
of a system of precision tracking chambers ($|\eta|<2.7$), and detectors for triggering ($|\eta|<2.4$).

\section{Monte Carlo simulation}

Monte Carlo (MC) simulated event samples are used to develop and validate the analysis procedure 
and to evaluate the subdominant SM backgrounds as well as the expected signal yields.
The dominant SM background processes include \ttbar, single-top, and diboson
($WW$, $WZ$ and $ZZ$) production.
The predictions for the most relevant SM processes are normalized to data in dedicated control regions, 
as detailed in section~\ref{sec:backgrounds}.
MC samples are produced using a
{\tt GEANT4}~\cite{geant4} based detector simulation~\cite{atlassimulation}
or a fast simulation using a parameterization of the performance of the
ATLAS electromagnetic and hadronic calorimeters~\cite{atlfast2,fastcalosim} 
and {\tt GEANT4} elsewhere. 
The effect of multiple proton-proton collisions from the same or different bunch crossings 
is incorporated into the simulation by overlaying minimum bias events generated 
using {\tt PYTHIA}~\cite{pythia} onto hard scatter events.
Simulated events are weighted to match the distribution of the number of interactions per bunch crossing observed in data.

Production of top-quark pairs is simulated at next-to-leading order (NLO) with \mcatnlo\ v4.06 \cite{mcatnlo1,mcatnlo2,mcatnlo3}, 
assuming a top-quark mass of $172.5$~GeV\@. Additional samples generated with \powheg-\texttt{BOX} v1.0 \cite{powheg} and 
\acer\ v3.8~\cite{acermc} are used for the evaluation of systematic uncertainties. 
The \ttbar\ cross-section is normalized to the next-to-next-to-leading order (NNLO) 
calculation including resummation of next-to-next-to-leading logarithmic (NNLL) 
soft gluon terms obtained with \texttt{Top++} v2.0~\cite{Czakon:2011xx}.
Single top production is modelled with \mcatnlo\ v4.06 for $Wt$ and $s$-channel production, and with \acer\ v3.8 for $t$-channel production.
Production of \ttbar\ associated with a vector boson is simulated with the leading-order (LO) generator \madgraph~5 v1.3.33 \cite{madgraph} 
and normalized to the NLO cross-section~\cite{Lazopoulos:2008de,Campbell:2012dh,Garzelli:2012bn}.

Diboson ($WW$, $WZ$ and $ZZ$) production is simulated with \powheg-\texttt{BOX} v1.0, with additional gluon-gluon contributions 
simulated with \texttt{gg2WW} v3.1.2 \cite{gg2WW} and \texttt{gg2ZZ} v3.1.2 \cite{gg2ZZ}. Additional diboson samples are generated 
at the particle level with \amcnlo\ v2.0 \cite{aMC@NLO} to assess systematic uncertainties.
The diboson cross-sections are normalized to NLO QCD predictions obtained with \mcfm\ v6.2 \cite{Campbell:1999ah,Campbell:2011bn}. 
Triple-boson ($WWW$, $ZWW$ and $ZZZ$) production is simulated with \madgraph~5 v1.3.33 \cite{tribosonSF},
and vector-boson scattering ($WWjj$ and $WZjj$) is simulated with \sherpa\ v1.4.1 \cite{sherpa}.

Samples of $W\to\ell\nu$ and $Z/\gamma^*\to\ell\ell$ produced with accompanying jets (including light and heavy flavours) are 
obtained with a combination of \sherpa\ v1.4.1 and \alpgen\ v2.14 \cite{alpgen}. The inclusive $W$ and $Z/\gamma^*$ production 
cross-sections are normalized to the NNLO cross-sections obtained using \dynnlo\ v1.1 \cite{Catani:2009sm}.
QCD production of \bbbar\ and \ccbar\ is simulated with \pythia\ v8.165.

Finally, production of the SM Higgs boson with $m_H=125\GeV$ is considered. The gluon fusion and vector-boson fusion production modes 
are simulated with \powheg-\texttt{BOX} v1.0, and the associated production ($WH$ and $ZH$) with \pythia\ v8.165.

Fragmentation and hadronization for the \mcatnlo\ and \alpgen\ samples are performed either with \herwig\ v6.520~\cite{herwig} using \jimmy\ v4.31~\cite{jimmy}
for the underlying event, or with \pythia\ v6.426. \pythia\ v6.426 is also used for \madgraph\ samples, whereas \pythia\ v8.165 is 
used for the \powheg-\texttt{BOX} samples. 
For the underlying event, ATLAS tune AUET2B~\cite{mc11tune} is used.
The CT10 NLO~\cite{CT10} and CTEQ6L1~\cite{CTEQ6} parton-distribution function (PDF) 
sets are used with the NLO and LO event generators, respectively.

Simulated signal samples are generated with \texttt{HERWIG++} v2.5.2~\cite{Bahr:2008pv}
and the CTEQ6L1 PDF set.
Signal cross-sections are calculated to NLO using \prospino2.1~\cite{prospinoCNSL}.
They are in agreement with the NLO calculations matched to resummation at the next-to-leading logarithmic accuracy (NLO+NLL) within $\sim2\%$~\cite{Fuks:2012qx,Fuks:2013vua,Fuks:2013lya}.

\section{Event reconstruction}
\label{sec:objectreconstruction} 

Events are selected in which at least five tracks,
each with transverse momentum $\pT>400\MeV$,
are associated to the primary vertex.
If there are multiple primary vertices in an event, the one with the largest $\sum\pt^2$ of the associated
tracks is chosen.
In each event, `candidate' electrons, muons, hadronically-decaying $\tau$ leptons, 
and jets are reconstructed.
After resolving potential ambiguities among objects, the criteria to
define `signal' electrons, muons and jets are refined.
Hadronically-decaying $\tau$ leptons are not considered as signal leptons for this analysis, 
and events containing them are removed (see section~\ref{sec:selection})
so that the data sample is distinct from that used in
the ATLAS search for electroweak SUSY production in the three-lepton 
final states~\cite{ATLAS_3LEWK2013}. 

Electron candidates are reconstructed by matching clusters in the EM calorimeter 
with tracks in the ID\@.
The magnitude of the momentum of the electron is determined by the calorimeter cluster energy.
They are required to have $\pt>10\GeV$, $|\eta|<2.47$, and satisfy shower-shape and
track-selection criteria analogous to the `medium' criteria in ref.~\cite{egammaperf}.

Muon candidates are reconstructed by matching an MS track to an ID track~\cite{muon2012}. 
They are then required to have $\pT>10\GeV$ and $|\eta| < 2.4$. 

Jet candidates are reconstructed from calorimeter energy clusters using the anti-$k_t$ jet clustering algorithm~\cite{anti-kt,anti-kt2} with a radius parameter of $0.4$. 
The jet candidates are corrected for the effects of calorimeter response and inhomogeneities  using energy- and $\eta$-dependent calibration factors based on simulation and validated with extensive test-beam and collision-data studies \cite{Aad:2011he}. 
Energy deposition due to pile-up interactions is statistically subtracted based on the
area of the jet~\cite{jetarea}.
Only jet candidates with $\pT>20\GeV$ and $|\eta|<4.5$ are subsequently retained.
Events containing jets that are likely to have arisen from detector noise or cosmic rays are removed~\cite{Aad:2011he}. 

A $b$-jet identification algorithm~\cite{btag7tev}
is used to identify jets containing a $b$-hadron decay inside a candidate jet
within $|\eta|<2.4$,
exploiting the long lifetime of $b$- and $c$-hadrons.
The mean nominal $b$-jet identification efficiency, determined from 
simulated $t\bar{t}$ events, 
is 80\%.
The misidentification (mis-tag) rates for $c$-jets and light-quark/gluon jets 
are approximately 30\% and 4\%, respectively.
Small differences in the $b$-tagging performance observed between data and simulation are
corrected for as functions of \pt\ of the jets.

Hadronically-decaying $\tau$ leptons are reconstructed by associating tracks with $\pt>1\GeV$ passing
minimum track quality requirements to calorimeter jets with $\pT>10\GeV$ and $|\eta|<2.5$.
A multivariate discriminant is used to identify the jets as hadronic $\tau$ decays~\cite{tauId2012}.
Their energy is determined by applying a simulation-based correction to the reconstructed energy 
in the calorimeter~\cite{tauECalibration}, and $\pt>20\GeV$ is required.

Object overlaps are defined in terms of $\Delta R=\sqrt{(\Delta\eta)^2+(\Delta\phi)^2}$,
where $\Delta\eta$ and $\Delta\phi$ are separations in $\eta$ and $\phi$.
Potential ambiguities among objects are resolved by removing 
one or both of nearby object pairs in the following order:
if two electron candidates are within $\Delta R=0.05$ of each other, the electron with
the smaller \pt\ is removed;
any jet within $\Delta R=0.2$ of an electron candidate is removed;
any $\tau$ candidate within $\Delta R=0.2$ of an electron or a muon is removed;
any electron or muon candidate within $\Delta R=0.4$ of a jet is removed;
if an electron candidate and a muon candidate are within $\Delta R=0.01$ of each other, both are removed;
if two muon candidates are within $\Delta R=0.05$ of each other, both are removed;
if the invariant mass of a SF opposite-sign lepton pair has an invariant mass less than $12\GeV$, both are removed;
and finally any jet within $\Delta R=0.2$ of a $\tau$ candidate is removed.

Signal electrons are electron candidates satisfying the `tight' criteria~\cite{egammaperf} 
placed on the ratio of calorimetric energy to track momentum, 
and the number of high-threshold hits in the transition radiation tracker.
They are also required to be isolated.
The $\pT$ scalar sum of tracks above $400\MeV$ within a cone of size $\Delta R=0.3$ 
around each electron candidate 
(excluding the electron candidate itself) and associated to the primary vertex is 
required to be less than 16\% of the electron $\pT$. 
The sum of transverse energies of the surrounding calorimeter clusters 
within $\Delta R =0.3$ of each electron candidate, 
corrected for the deposition of energy from pile-up interactions, 
is required to be less than 18\% of the electron \pt.
The distance of closest approach of an electron candidate to the event primary vertex 
must be within five standard deviations in the transverse plane.
The distance along the beam direction, $z_0$, must satisfy $|z_0\sin\theta|<0.4$\,mm.

Signal muons are muon candidates satisfying the following criteria. 
The $\pT$ scalar sum of tracks above $400\MeV$ within a cone of size $\Delta R=0.3$ 
around the muon candidate and associated to the primary vertex is required to be 
less than 16\% of the muon $\pT$.
The distance of closest approach of a muon candidate to the event primary vertex must be 
within three standard deviations in the transverse plane, 
and $|z_0\sin\theta|<1\,\mathrm{mm}$ along the beam direction.

The efficiencies for electrons and muons to pass the reconstruction, identification and
isolation criteria are measured in samples of $Z$ and $J/\psi$
leptonic decays, and corrections are applied to the simulated samples to reproduce the
efficiencies in data.

Signal jets are jet candidates that are classified in three exclusive categories.
Central $b$-jets satisfy $|\eta|<2.4$ and the $b$-jet identification criteria.
Central light-flavour jets also satisfy $|\eta|<2.4$ but do not satisfy the $b$-jet identification criteria.
If a central light-flavour jet has $\pT<50\GeV$ and has tracks associated to it,
at least one of the tracks must originate from the event primary vertex.
This criterion removes jets that originate from pile-up interactions.
Finally, forward jets are those with $2.4<|\eta|<4.5$ and $\pT>30\GeV$.

The missing transverse momentum, \pTmiss, is defined~\cite{atlasMET}
as the negative vector sum 
of the total transverse momenta of all $\pT>10\GeV$ electron, muon and photon candidates, 
$\pT>20\GeV$ jets, and all clusters of calorimeter 
energy with $|\eta|<4.9$ not associated to such objects, referred to hereafter as the `soft-term'. Clusters associated with
electrons, photons and jets make use of calibrations of the respective objects, whereas clusters not associated 
with these objects are calibrated using both calorimeter and tracker information.

The quantity \METrel\ is defined from the magnitude, \met, of \pTmiss\ as 
\begin{equation}
	\METrel =  \left\{   
	\begin{array}{ll}  
	   \met & \quad \text{if $\Delta \phi_{\ell,j} \geq \pi$/2} \\
	   \met\times\sin{\Delta \phi_{\ell,j}} & \quad \text{if  $\Delta \phi_{\ell,j} < \pi$/2} \\
    \end{array} \right., \nonumber
\end{equation}
where $\Delta \phi_{\ell,j}$ is the azimuthal angle between the direction of 
\pTmiss\ and that of the nearest  electron,  muon, central $b$-jet
or central light-flavour jet. 
Selections based on \METrel\ aim to suppress events where missing
transverse momentum arises from significantly mis-measured jets or leptons.

\section{Event selection}\label{sec:selection}

Events are recorded using a combination of two-lepton triggers, 
which require identification of two
lepton (electron or muon) candidates with transverse momenta exceeding a set of thresholds.
For all triggers used in this measurement, 
the \pt\ thresholds are 18--$25\GeV$ for the higher-\pt\ lepton
and 8--$14\GeV$ for the other lepton.
After event reconstruction, two signal leptons of opposite charge,
with $\pt>35\GeV$ and $>20\GeV$, are required in the selected events.
No lepton candidates other than the two signal leptons are allowed in the event.
The two signal leptons are required to match those that triggered the event.
The trigger efficiencies 
with respect to reconstructed leptons with $\pt$ in excess of the nominal thresholds
have been measured using data-driven techniques.
For events containing two reconstructed signal leptons with $\pt>35\GeV$ and $>20\GeV$, the average trigger efficiencies are 
approximately 97\% in the \ee\ channel, 75\% in the \emu\ channels, 
and 89\% in the \mumu\ channel.

The dilepton invariant mass \mll\ must be greater than $20\GeV$ in all flavour combinations.
Events containing one or more $\tau$-jet candidates are rejected.

Seven signal regions (SRs) are defined in this analysis.
The first three, collectively referred to as \SRmttwo, are designed to provide sensitivity to sleptons
either through direct production or in chargino decays.
The next three, \SRWW, are designed to provide sensitivity to chargino-pair production
followed by $W$ decays.
The last signal region, \SRZjets, is designed specifically 
for chargino and second lightest neutralino associated production followed by 
hadronic $W$ and leptonic $Z$ decays.
The SF and DF event samples in each SR are considered separately.
When a scenario that contributes to both SF and DF final states is considered,
a simultaneous fit to the SF and DF samples is employed.
All SRs of the same lepton flavour combination,
except for \SRZjets, overlap with each other and are not statistically independent.
Table~\ref{SR} summarizes the definitions of the SRs.

Five of the SRs exploit the `stransverse' mass \mttwo~\cite{Lester:1999tx,Barr:2003rg}, defined as
\newcommand{\pTvec}{\mathbf{p}_\mathrm{T}}
\newcommand{\qTvec}{\mathbf{q}_\mathrm{T}}
\newcommand{\qT}{q_\mathrm{T}}
\newcommand{\mT}{m_\mathrm{T}}
\newcommand{\pTell}[1]{\mathbf{p}_\mathrm{T}^{\ell#1}}
\[
\mttwo = \min_{\qTvec}\left[\max\left(\mT(\pTell{1},\qTvec),\mT(\pTell{2},\pTmiss-\qTvec)\right)\right],
\]
where $\pTell{1}$ and $\pTell{2}$ are the transverse momenta of the two leptons,
and $\qTvec$ is a transverse vector that minimizes the larger of the
two transverse masses $\mT$.
The latter is defined by
\[
\mT(\pTvec,\qTvec) = \sqrt{2(\pT\qT-\pTvec\cdot\qTvec)}.
\]
For SM \ttbar\ and $WW$ events, in which two $W$ bosons decay leptonically and $\pTmiss$
originates from the two neutrinos, the \mttwo\ distribution has an upper end-point at the $W$ mass.
For signal events, the undetected LSP contributes to $\pTmiss$, and
the \mttwo\ end-point is correlated to the mass difference between
the slepton or chargino and the lightest neutralino. 
For large values of this difference, the \mttwo\
distribution for signal events extends significantly beyond the distributions of the  \ttbar\ and $WW$ events.

\subsection{\SRmttwo}

\SRmttwo\ targets \ConeCone\ production followed by slepton-mediated decays
(figure~\ref{diagrams}(a)) and direct slepton pair production
(figure~\ref{diagrams}(d)).
Events are required to contain two opposite-sign signal leptons and no signal jets.
Only SF channels
are used in the search for direct slepton production, while the chargino-to-slepton decay
search also uses DF channels.
In the SF channels, the dilepton invariant mass \mll\ must be at least
$10\GeV$ away from the $Z$ boson mass.

The dominant sources of background are diboson and top production ($t\bar{t}$ and $Wt$).
Three signal regions, \SRmtt{90}, \SRmtt{120} and \SRmtt{150}, are defined by requiring
$\mttwo>90\GeV$, $120\GeV$ and $150\GeV$, respectively. 
Low values of \mttwo\ threshold provide
better sensitivity to cases in which the \slepton\ or \Cone\ mass is close
to the \None\ mass, and high values target large \slepton--\None\ or \Cone--\None\
mass differences. 

\subsection{\SRWW}

Direct \ConeCone\ production followed by $W$-mediated decays 
(figure~\ref{diagrams}(b)) is similar to the
slepton-mediated scenario, but with smaller visible cross-sections due to the
$W\to\ell\nu$ branching fraction.
Three signal regions, \SRWWa, \SRWWb\ and \SRWWc, are designed to provide
sensitivities to this scenario for increasing values of \chargino{1}--\neutralino{1} mass difference.
Events are required to contain two opposite-sign signal leptons and no signal jets.
Both SF and DF channels are used in these signal regions.
In the SF channels, the dilepton invariant mass \mll\ must be at least
$10\GeV$ away from the $Z$ boson mass.

For large \chargino{1}--\neutralino{1} mass splitting,
the \mttwo\ variable provides good discrimination between the signal and SM background.
Two signal regions, \SRWWb\ and \SRWWc, are defined by
$\mttwo>90\GeV$ and $100\GeV$, respectively.
The \mttwo\ thresholds are lower than in \SRmttwo\ because the smaller visible cross-sections 
limit the sensitivity to large \chargino{1} masses.
For \SRWWb, an additional requirement of $\mll<170\GeV$ is applied to further suppress the 
SM background.

For cases in which the \chargino{1}--\neutralino{1} mass splitting is close to the $W$ boson mass,
the \mttwo\ variable is not effective in distinguishing signal from the SM $WW$ production.
The signal region \SRWWa\ is defined by 
$\METrel>80\GeV$, $\pTll>80\GeV$ and $\mll<120\GeV$,
where \pTll\ is the transverse momentum of the lepton pair. 
These selection criteria favour events in which the di-lepton opening angle is small, 
which enhances the difference in the \METrel\ distribution 
between the signal and the background due to the two LSPs in the signal.

\subsection{\SRZjets}

The last signal region, \SRZjets, differs from the previous six in that it requires the
presence of at least two central light jets.
This signal region is designed to target
the $pp\to\chargino{1}\neutralino{2}\to W^\pm\neutralino{1}Z\neutralino{1}$ process
in which the $W$ boson decays hadronically and the $Z$ boson decays leptonically
(figure~\ref{diagrams}(c)).

The two highest-\pT\ central light jets must have $\pT>45\GeV$, and
have an invariant mass in the range $50<\mjj<100\GeV$.
There must be no central $b$-jet and no forward jet in the event.
The two opposite-sign leptons must be SF, and their invariant mass
must be within $10\GeV$ of the $Z$ boson mass.

To suppress large background from the SM $Z+\mathrm{jets}$ production,
$\METrel>80\GeV$ is required.
Events are accepted only if the reconstructed $Z$ boson is recoiling against the 
rest of the event with a large transverse momentum $\pTll>80\GeV$, 
and the separation $\dRll$ between the two leptons must satisfy
$0.3<\dRll<1.5$.

\begin{table}
\begin{center}
\caption{\label{SR}
	Signal region definitions.
	The criteria on $|\mll-m_Z|$ are applied only to SF events.
	The two leading central light jets in \SRZjets\ must have $\pT>45\GeV$.
}\smallskip
\begin{tabular}{r|ccc|ccc|c}
\hline
 SR & $m_{\mathrm{T2}}^{90}$ & $m_{\mathrm{T2}}^{120}$ & $m_{\mathrm{T2}}^{150}$ 
    & $WW$a & $WW$b & $WW$c & $Z$jets \\
\hline 
 lepton flavour     & DF,SF & DF,SF  & DF,SF   & DF,SF  & DF,SF  & DF,SF  & SF \\
 central light jets & 0     & 0      & 0       & 0      & 0      & 0      & $\ge2$ \\
 central $b$-jets   & 0     & 0      & 0       & 0      & 0      & 0      & 0 \\
 forward jets       & 0     & 0      & 0       & 0      & 0      & 0      & 0 \\
 $|\mll-m_Z|$ [GeV] & $>10$ & $>10$  & $>10$   & $>10$  & $>10$  & $>10$  & $<10$ \\
 $\mll$ [GeV]       & ---   & ---    & ---     & $<120$ & $<170$ & ---    & --- \\
 \METrel\ [GeV]     & ---   & ---    & ---     & $>80$  & ---    & ---    & $>80$ \\
 \pTll\ [GeV]       & ---   & ---    & ---     & $>80$  & ---    & ---    & $>80$ \\
 \mttwo\ [GeV]      & $>90$ & $>120$ & $>150$  &  ---   & $>90$  & $>100$ & --- \\
 \dRll\             & ---   & ---    & ---     &  ---   & ---    & ---    & [0.3,1.5] \\
 \mjj\ [GeV]        & ---   & ---    & ---     &  ---   & ---    & ---    & [50,100] \\
\hline
\end{tabular}
\end{center}
\end{table}

\section{Background estimation}\label{sec:backgrounds}

For \SRmttwo\ and \SRWW, 
the SM background is dominated by $WW$ diboson and top-quark (\ttbar\ and $Wt$) production.
Contributions from $ZV$ production, where $V=W$ or $Z$, are also significant
in the SF channels.
The MC predictions for these background sources are normalized
in dedicated control regions (CR) for each background, 
as described in section~\ref{sec:mttWW}.
For \SRZjets, the dominant sources of background are $ZV$ production and $Z/\gamma^*+\mathrm{jets}$.
The former is estimated from simulation,
validated using $ZV$-enriched control samples,
and the latter is estimated by a data-driven technique,
as described in section~\ref{sec:Zjets}.
The top-quark background in \SRZjets\ is estimated using a dedicated CR\@.
Background due to hadronic jets mistakenly reconstructed as signal leptons
or real leptons originating from heavy-flavour decays or photon conversions,
referred to as `\fake\ leptons', is estimated using a data-driven method
described in section~\ref{sec:fake}.
Contributions from remaining sources of SM background are small and are estimated from  simulation.
Table~\ref{tab:CRdef} summarizes the definitions of the control regions.

\subsection{Background in \SRmttwo\ and \SRWW}\label{sec:mttWW}

The normalization factors for the background in \SRmttwo\ and \SRWW\ due to the SM $WW$, top and $ZV$ production 
are constrained in dedicated CRs for each background. 
Each CR is dominated by the background of interest and is designed to be 
kinematically as close as possible to a corresponding signal region.
The normalization factors are obtained from the likelihood
fit described in section~\ref{sec:fit}.

\begin{table}
\begin{center}
\caption{\label{tab:CRdef}
	Control region definitions.
	The top CR for \SRZjets\ requires at least two jets with $\pt>20\GeV$ in
	$|\eta|<2.4$, at least one of which is $b$-tagged.
}\smallskip
\begin{tabular}{r|ccc|ccc|c}
\hline
SR & \multicolumn{3}{c|}{\mttwo\ and $WW$b/c} & \multicolumn{3}{c|}{$WW$a} & $Z$jets \\
\hline 
CR                 & $WW$      & Top    & $ZV$  & $WW$      & Top    & $ZV$  & Top         \\
\hline
lepton flavour     & DF        & DF     & SF    & DF        & DF     & SF    & SF          \\
central light jets & 0         & 0      & 0     & 0         & 0      & 0     & $\ge2$    \\
central $b$-jets   & 0         & $\ge1$ & 0     & 0         & $\ge1$ & 0     & $\ge1$    \\
forward jets       & 0         & 0      & 0     & 0         & 0      & 0     & 0           \\
$|\mll-m_Z|$ [GeV] & ---       & ---    & $<10$ & ---       & ---    & $<10$ & $>10$       \\
\mll\ [GeV]        & ---       & ---    & ---   & $<120$    & $<120$ & ---   & ---         \\
\METrel\ [GeV]     & ---       & ---    & ---   & $[60,80]$ & $>80$  & $>80$ & $>80$       \\
\pTll\ [GeV]       & ---       & ---    & ---   & $>40$     & $>80$  & $>80$ & $>80$       \\
\mttwo\ [GeV]      & $[50,90]$ & $>70$  & $>90$ & ---       & ---    & ---   & ---         \\
\dRll\             & ---       & ---    & ---   & ---       & ---    & ---   & $[0.3,1.5]$ \\
\hline
\end{tabular}
\end{center}
\end{table}

\begin{figure}
\centering
    \raisebox{0.38\textwidth}{(a)}\plot{0.45}{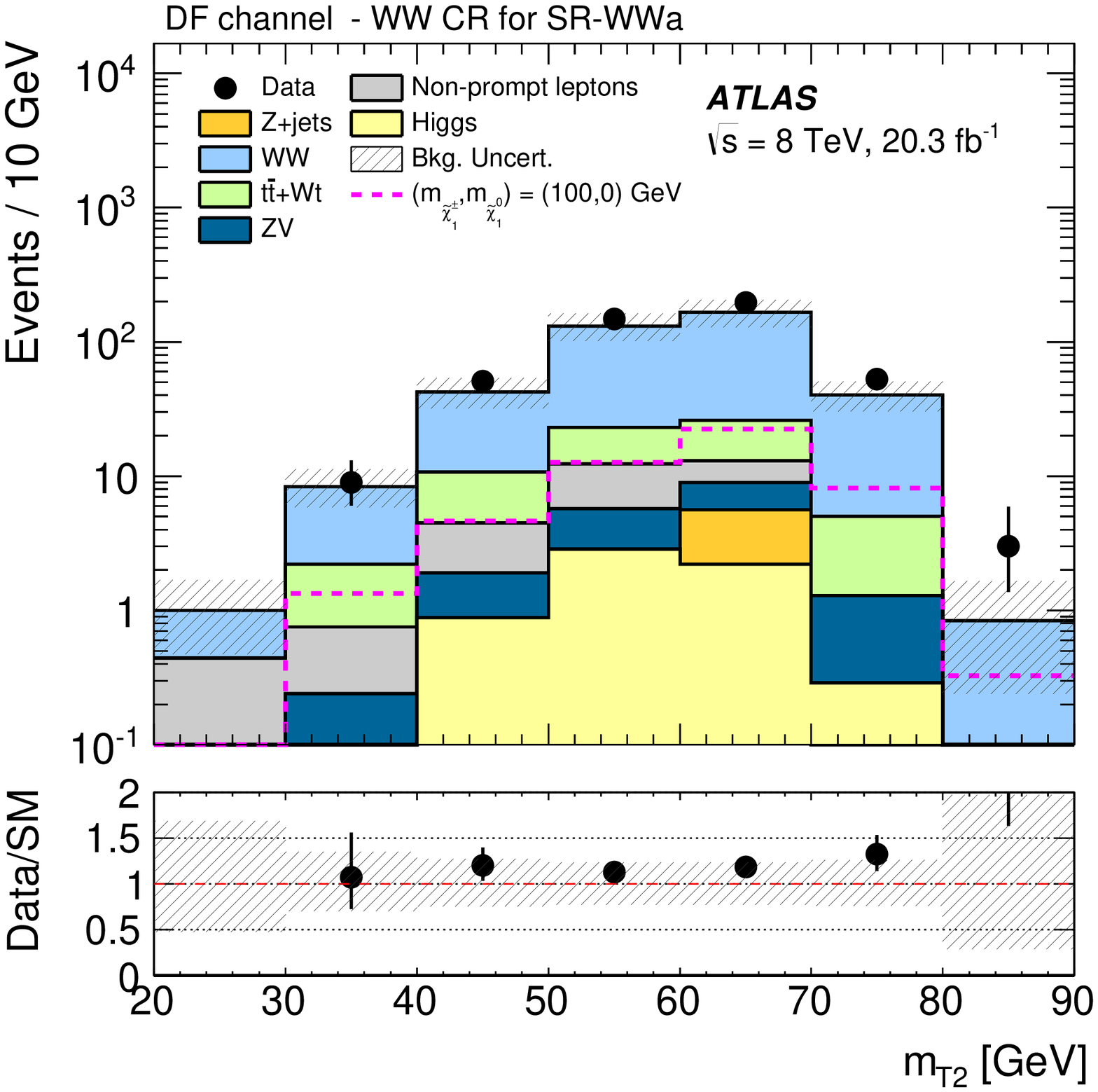}{74,86}
    \raisebox{0.38\textwidth}{(b)}\plot{0.45}{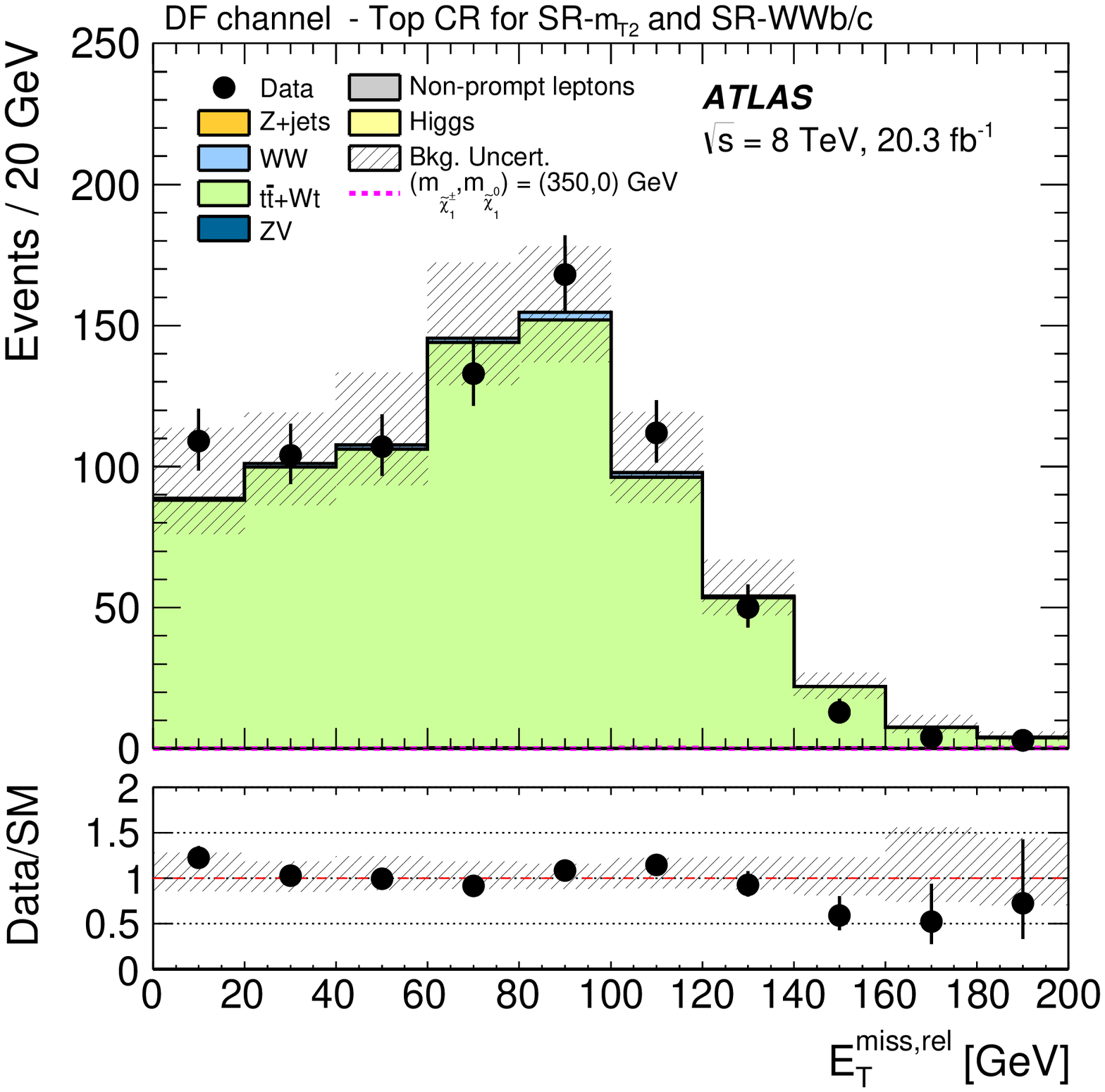}{74,86}
    \raisebox{0.38\textwidth}{(c)}\plot{0.45}{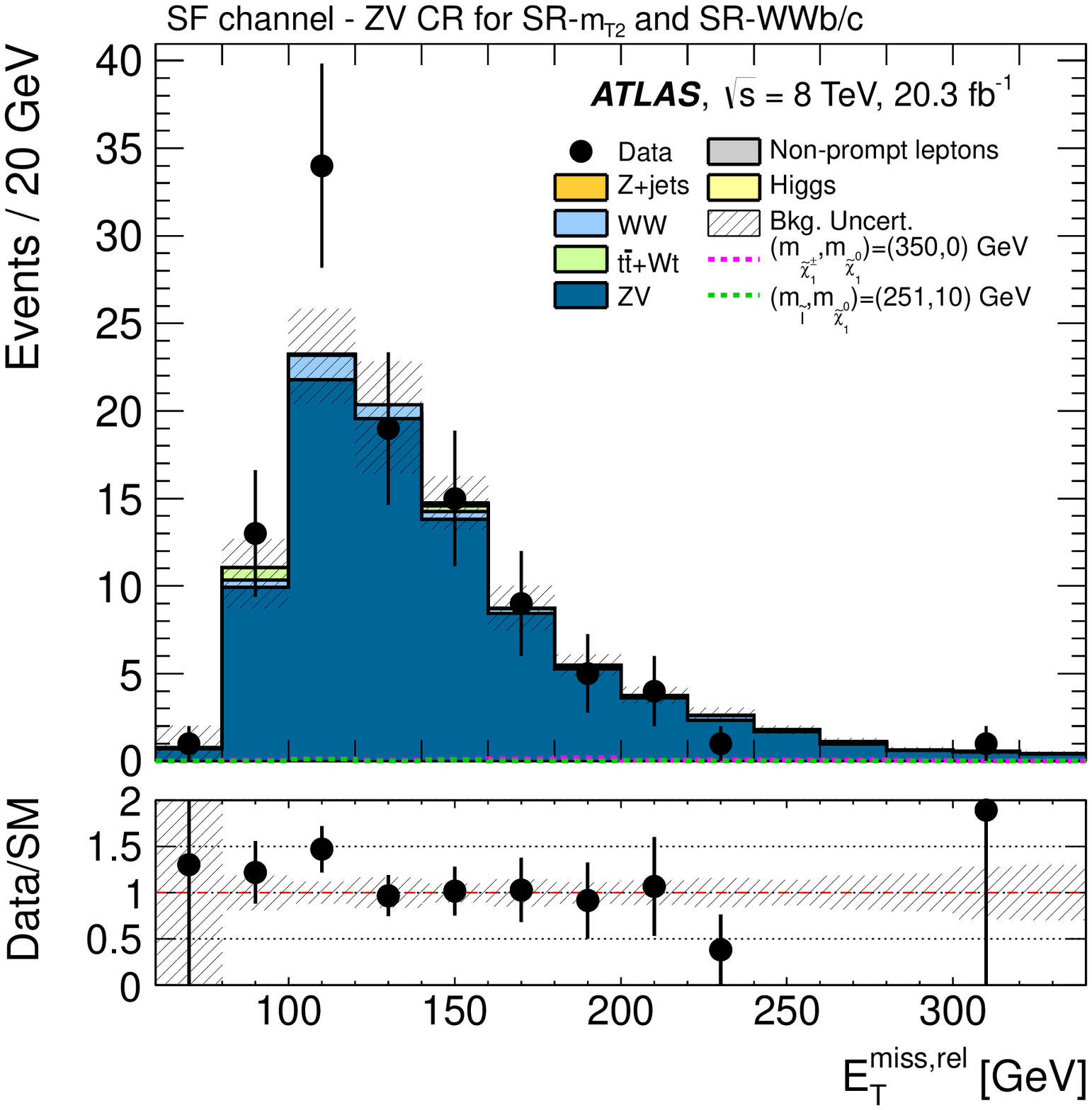}{40,86}
    \raisebox{0.38\textwidth}{(d)}\plot{0.45}{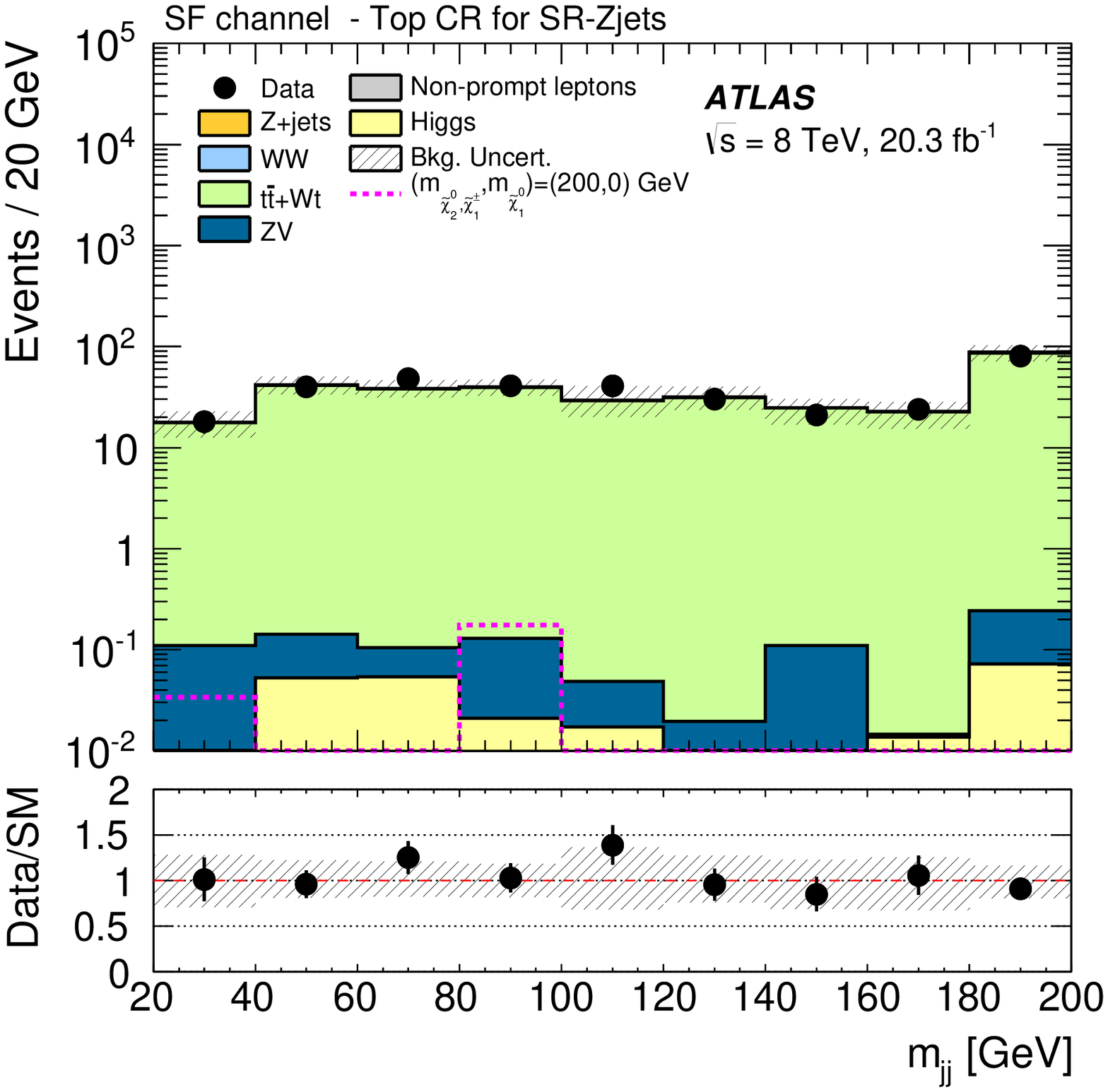}{74,86}
	\caption{\label{fig:CRplots}
		Distributions of
		(a) \mttwo\ in the $WW$ CR for \SRWWa,
	    (b) \METrel\ in the top CR for \SRWWb/c and \SRmttwo,
        (c) \METrel\ in the $ZV$ CR for \SRWWb/c and \SRmttwo, and
        (d) $m_{jj}$ in the top CR for \SRZjets.
		No data-driven normalization factor is applied to the distributions.
		The hashed regions represent the total uncertainties on the background estimates.
		The rightmost bin of each plot includes overflow.
		The lower panel of each plot shows the ratio between data and the 
		SM background prediction.
	}
\end{figure}

The $WW$ control region for \SRmttwo\ and \SRWWb/c is defined by requiring $50<\mttwo<90\GeV$
and the events must contain no jets.
Only the DF sample is used in this CR because the corresponding regions
in the SF samples suffer from contamination from $Z/\gamma^*+\mathrm{jets}$ background.
Appropriate ratios of electron and muon efficiencies are used to obtain 
the SF background estimations from the corresponding DF CR\@. 
For SR-$WW$a, the CR is defined by lowering the \METrel\ and \pTll\ requirements 
so that $60<\METrel<80\GeV$ and $\pTll>40\GeV$. 
Figure~\ref{fig:CRplots}(a) shows the \mttwo\ distribution in this CR\@.
The normalization factors are not applied to the MC predictions in all four
plots of figure~\ref{fig:CRplots}.
Predicted signal contamination in this CR is less than 10\%
for the signal models 
$\tilde{\chi}^\pm_1\tilde{\chi}^\mp_1 \rightarrow W^{\pm}W^{\mp}\neutralino{1}\neutralino{1}$ 
with $m_{\chargino{1}}>100$~\GeV.

The top control region for \SRmttwo\ and \SRWWb/c
is also defined using the DF sample, and by requiring
at least one $b$-tagged jet and vetoing central light jets and forward jets.
The events must also satisfy $\mttwo>70\GeV$.
Figure~\ref{fig:CRplots}(b) shows the \METrel\ distribution in this CR\@.
For \SRWWa, the CR is defined using the DF sample and requiring at least one $b$-tagged jet, 
with all the other SR criteria unchanged.
The predicted contamination from SUSY signal is negligible for the models considered.

The $ZV$ control region for \SRmttwo\ and \SRWWb/c is defined identically to the SF \SRmtta, 
but with the $Z$ veto reversed.
Figure~\ref{fig:CRplots}(c) shows the \METrel\ distribution in this CR\@.
The contamination due to non-$ZV$ sources is dominated by $WW$ events (4.5\%).
For \SRWWa, the CR is defined by reversing the $Z$ veto in the SF sample. 
The predicted contamination from SUSY signal is less than 5\% in these CRs.

\subsection{Background in \SRZjets}\label{sec:Zjets}

The top CR for \SRZjets\ is defined by reversing the $Z$ veto
and requiring at least one $b$-tagged jet.
To increase the statistics of the sample, 
the $\pT$ threshold for the central jets is lowered to $20\GeV$, 
and no cut on $\mjj$ is applied.
Figure~\ref{fig:CRplots}(d) shows the $m_{jj}$ distribution in this CR\@.
The predicted contamination from SUSY signal is negligible.

The $ZV$ background in \SRZjets\ consists of diboson production accompanied by 
two light-flavour jets, that is, $WZjj\to\ell\nu\ell'\ell'jj$, where
the lepton from the $W$ decay was not reconstructed, and $ZZjj\to\ell\ell\nu\nu jj$.
The contribution from $ZV\to\ell\ell\qqbar$ is strongly suppressed by the \METrel\ requirement.
This background is estimated from simulation,
and validated in control samples of $WZjj\to\ell\nu\ell'\ell'jj$ and $ZZjj\to\ell\ell\ell'\ell'jj$
where all leptons are reconstructed. 
The $WZjj$ enriched control sample consists of events with three leptons,
at least two of which make up a SF opposite-sign pair with an invariant mass
within $10\GeV$ of the $Z$ boson mass.
In addition, events must have $\MET>30\GeV$, $\mT>40\GeV$ computed from the \pTmiss\ and the
lepton that was not assigned to the $Z$ boson, at least two central light jets,
and no central $b$-jet. 
The predicted contamination from SUSY signal is less than 10\% in this region.
The $ZZjj$ enriched control sample consists of events with two pairs of
same-flavour opposite-sign leptons,
 each with an invariant mass within $10\GeV$ of the $Z$ boson mass, 
$\MET<50\GeV$, at least two central light jets, and no signal $b$-jet. 
The data in these control samples are compared with the simulation to assess the
systematic uncertainties of the $ZV$ background estimation, as reported in section~\ref{sec:Systematics}.

In \SRZjets,  $Z/\gamma^*+\mathrm{jets}$ events are an important source of background, 
where significant \MET\ arises primarily from mis-measurement of jet transverse momentum. 
A data-driven approach called the `jet smearing' method 
is used to estimate this background.
In this method, a sample enriched in $Z/\gamma^*+\mathrm{jets}$ events with well-measured 
jets is selected from data as seed events.
The seed events are selected by applying the \SRZjets\ event selection, 
but reversing the \METrel\ cut. 
To ensure that the events only contain well measured jets,
the ratio $\MET/\sqrt{\ET^\mathrm{sum}}$,
where $\ET^\mathrm{sum}$ is the scalar sum of the transverse energies of the jets and the soft-term,
is required to be less than $1.5\,\mathrm{(GeV)}^{1/2}$.
Each seed event is smeared by multiplying each jet four-momentum by a random number 
drawn from the jet response function,
which is initially estimated from simulation and adjusted after comparing the response 
to data in a photon\,+\,jet sample.
In addition, the contribution to \MET\ due to the soft-term
is also modified by sampling randomly from the soft-term 
distribution measured in a $Z\to\ell\ell$ sample with no reconstructed jets.
The smearing procedure is repeated 10,000 times for each seed event.
The resulting pseudo-data \METrel\ distribution is then normalized to the data in 
the region of $\METrel<40\GeV$, and the migration into the signal region is evaluated.

To validate the jet-smearing method, a control sample is selected with the same selection criteria as \SRZjets\ 
but reversing the \pTll\ requirement, and removing the \dRll\ and \mjj\ criteria to increase the 
number of events. 
The seed events are selected from the control region events by requiring 
$\METrel<40\GeV$ and $\MET/\sqrt{\ET^\mathrm{sum}}<1.5\,\mathrm{(GeV)}^{1/2}$. 
Results are validated in a region with $40<\METrel<80\GeV$, 
which is dominated by $Z/\gamma^*+\mathrm{jets}$.
The method predicts $750\pm100$ events, where both statistical and systematic uncertainties are included,
in agreement with the 779 events observed in data.

\subsection{\Fake\ lepton background estimation}\label{sec:fake}

The term `\fake\ leptons' refers to hadronic jets mistakenly reconstructed as signal leptons
or leptons originating from heavy-flavour decays or photon conversions.
In this context, `\real\ leptons' are leptons produced directly in decays of sparticles or weak bosons.
The number of \fake\ lepton events is estimated using the matrix method~\cite{atlas-topxsec},
which takes advantage of the difference between the \real\ efficiency \reff\ and \fake\ efficiency \feff,
defined as the fractions of \real\ and \fake\ candidate leptons, 
respectively, that pass the signal-lepton requirements.

The \real\ and \fake\ efficiencies are evaluated as functions of the \pT\ of the lepton candidate
in simulated events using MC truth information.
Differences between data and MC are corrected for with normalization factors measured in  control samples.
Since the efficiencies depend on the production process, average \reff\ and \feff\ values
are calculated 
for each SR and CR using the fraction of each process predicted by the simulation as the weights.
The data/MC normalization factors for \reff\ are derived from $Z\to\ell\ell$ events.
The normalization factors for \feff\ depend on whether the \fake\ lepton originated from jets or 
from photon conversion.
The normalization factors for misidentified jets or leptons from
heavy-flavour decays are measured in a control region enriched in \bbbar\
production. Events are selected with two candidate leptons, one $b$-tagged jet and $\METrel<40\GeV$.
One of the two lepton candidates is required to be a muon and to lie within $\Delta R=0.4$ of the $b$-tagged jet,
while the other lepton candidate is used to measure the \fake\ efficiency.
For measuring the normalization factor for photon conversions, a $Z\to\mu\mu\gamma$ control sample is defined by selecting events
with two muons, $\METrel<50\GeV$, at least one candidate electron
(which is the conversion candidate) with $\mT<40\GeV$, 
and requiring that the invariant mass of the $\mumu e^{\pm}$ system is within $10\GeV$ of the $Z$ boson mass.

Using \feff\ and \reff, the observed numbers of events in each SR and CR with four possible 
combinations (signal-signal, signal-candidate, candidate-signal and 
candidate-candidate) of leptons are expressed as weighted sums of the numbers 
of events with four combinations of \real\ and \fake\ leptons.  
Solving these equations allows determination of the \fake\ lepton background. 
The contribution of \fake-lepton background in the signal regions is less than 5\% of the total background in all signal regions.

\subsection{Fitting procedure}\label{sec:fit}

For each SR, a simultaneous likelihood fit 
to the corresponding CRs is performed to normalize the top, $WW$ and $ZV$
(in the case of \SRZjets\ only top is fitted) background estimates.
The inputs to the fit are the numbers of observed events in the CRs,
the expected contributions of top, $WW$ and $ZV$ from simulation,
and the expected contributions of other background sources determined as
described in sections~\ref{sec:mttWW}--\ref{sec:fake}.

The event count in each CR is treated as a Poisson probability function,
the mean of which is the sum of the expected contributions from all background sources.
The free parameters in the fit are 
the normalization  of the top, $WW$ and $ZV$ contributions.
The systematic uncertainties on the expected background yields are included as
nuisance parameters, constrained to be Gaussian with a width determined 
from the size of the
uncertainty.
Correlations between control and signal regions, and background
processes are taken into account with common nuisance parameters.
The free parameters and the nuisance parameters are determined by maximizing 
the product of the Poisson probability functions and the
constraints on the nuisance parameters. 

\begin{table}
\centering
\caption{\label{tab:CRfit}
Numbers of observed and predicted events in the CRs, 
data/MC normalization factors and composition of the CRs obtained from the fit.
Systematic errors are described in section~\ref{sec:Systematics}.}\smallskip
\begin{tabular}{l|rrr|rrr|r}
\hline
SR & \multicolumn{3}{c|}{\mttwo\ and $WW$b/c} & \multicolumn{3}{c|}{$WW$a} & $Z$jets \\
\hline 
CR                  & $WW$   & Top    & $ZV$   & $WW$   & Top    & $ZV$   & Top    \\
\hline
Observed events     & 1061   & 804    & 94     & 472    & 209    & 175    & 395    \\
MC prediction       &  947   & 789    & 91     & 385    & 215    & 162    & 399    \\
\hline
Normalization       & 1.14   & 1.02   & 1.08   & 1.12   & 0.97   & 1.04   & 0.99   \\
Statistical error   & 0.05   & 0.04   & 0.12   & 0.08   & 0.08   & 0.12   & 0.06   \\
\hline
Composition         &        &        &        &        &        &        &        \\
\quad $WW$          & 84.6\% &  1.4\% &  5.0\% & 86.8\% &  1.7\% & 10.5\% &  1.3\% \\
\quad Top           & 10.4\% & 98.5\% &$<$0.1\%&  7.3\% & 98.1\% &  2.8\% & 98.0\% \\
\quad $ZV$          &  2.0\% &  0.1\% & 94.9\% &  1.9\% &$<$0.1\%& 82.9\% &  0.3\% \\
\quad \Fake\ lepton &  1.9\% &$<$0.1\%&$<$0.1\%&  2.7\% &$<$0.1\%&$<$0.1\%&$<$0.1\%\\
\quad Other         &  1.1\% &$<$0.1\%&  0.1\% &  1.3\% &$<$0.1\%&  3.7\% &  0.3\% \\
\hline
\end{tabular}
\end{table}
Table~\ref{tab:CRfit} summarizes the numbers of observed and predicted events in the CRs, 
data/MC normalization and CR composition obtained from the simultaneous fit.
The normalization factors agree within errors between different SRs for each
of the $WW$, Top and $ZV$ contributions.
Results of the background estimates in the SRs can be found in 
Tables~\ref{tab:expmt2}, \ref{tab:expww} and~\ref{tab:expzjets}.

\section{Systematic uncertainties}
\label{sec:Systematics}

Systematic uncertainties affect the estimates of the backgrounds and signal event yields
in the control and signal regions. 
The relative sizes of the sources of systematic uncertainty on the total SM background
in \SRmttwo, \SRWW and \SRZjets\ are detailed in Table~\ref{sys_summary}.

\begin{table}
\small
\begin{center}
\caption{\label{sys_summary}
Systematic uncertainties (in \%) on the total background estimated in different signal regions.
Because of correlations between the systematic uncertainties and the fitted backgrounds, the total
uncertainty can be different from the quadratic sum of the individual uncertainties.
}\smallskip
\begin{tabular}{l|rr|rr|rr|rr|rr|rr|rr}
\hline
& \multicolumn{2}{c|}{$\mttwo^{90}$} & \multicolumn{2}{c|}{$\mttwo^{120}$} & \multicolumn{2}{c|}{$\mttwo^{150}$}
& \multicolumn{2}{c|}{$WW$a} & \multicolumn{2}{c|}{$WW$b} & \multicolumn{2}{c|}{$WW$c} 
& \multicolumn{1}{c}{$Z$jets} \\
& SF & DF & SF & DF & SF & DF & SF & DF & SF & DF & SF & DF & SF \\
\hline
CR statistics       &    5 &    3 &  6 &  4 &  8 &  4 &  5 &  5 &  5 &  3 &  6 &  4 &  1 \\
MC statistics       &    5 &    7 &  7 & 12 & 10 & 23 &  3 &  4 &  5 &  8 &  6 & 10 & 14 \\
Jet                 &    4 &    1 &  2 &  1 &  5 &  7 &  3 &  6 &  4 &  2 &  4 &  3 & 11 \\
Lepton              &    1 &    2 &  1 &  1 &  4 &  1 &  1 &  3 &  2 &  3 &  1 &  8 &  4 \\
Soft-term           &    3 &    4 &  1 &  1 &  2 &  8 &$<1$&  2 &  3 &  5 &  1 &  6 &  5 \\
b-tagging           &    1 &    2 &$<$1&$<$1&$<$1&$<$1&  1 &  1 &  1 &  2 &$<$1&  1 &  2 \\
\Fake\ lepton       & $<$1 &    1 &$<$1&$<$1&  1 &$<$1&  1 &  1 &  1 &  2 &$<$1&  1 &$<$1\\
Luminosity          & $<$1 & $<$1 &$<$1&$<$1&$<$1&$<$1&$<$1&$<$1&$<$1&$<$1&$<$1&$<$1&  2 \\
Modelling           &   11 &   13 & 21 & 31 & 18 & 40 &  6 &  6 &  8 & 10 & 15 & 19 & 42 \\
\hline
Total               &   13 &   16 & 24 & 34 & 23 & 47 &  9 & 11 & 12 & 14 & 17 & 24 & 47 \\
\hline
\end{tabular}
\end{center}
\end{table}

The `CR statistics' and `MC statistics' uncertainties arise from the number of data events
in the CRs and simulated events in the SRs and CRs, respectively.
The largest contributions are due to the simulated background samples in the signal regions.

The dominant experimental systematic uncertainties, labelled `Jet' in Table~\ref{sys_summary},
come from the propagation of the jet energy scale calibration~\cite{JES1} 
and resolution~\cite{JER1} uncertainties.
They were derived from a combination of simulation, test-beam data and in situ measurements.
Additional uncertainties due to  differences between quark and gluon jets, and light and heavy
flavour jets, as well as the effect of pile-up interactions are included.
The `Lepton' uncertainties include those from lepton reconstruction, identification and trigger efficiencies,
as well as lepton energy and momentum measurements~\cite{egammaperf,muon2012}.
Uncertainties due to $\tau$ reconstruction and energy calibration are negligible.
Jet and lepton energy scale uncertainties are propagated to the \MET\ evaluation.
An additional `Soft-term' uncertainty is associated 
with the contribution to the \MET\ reconstruction of energy deposits not assigned to 
any reconstructed objects~\cite{atlasMET}.

The `$b$-tagging' row refers to the uncertainties on the $b$-jet identification efficiency and
charm and light-flavour jet rejection factors~\cite{bperf1}.
The `\Fake\ lepton' uncertainties arise from the data-driven estimates of the \fake\ lepton
background described in section~\ref{sec:fake}.  The dominant sources are 
$\eta$ dependencies of the \fake\ rates, differences between the light and heavy flavour jets,
and the statistics of the control samples.
The uncertainty on the integrated luminosity is $\pm\lumierror$, 
and affects the normalization of the background estimated with simulation. 
It is derived following the methodology detailed in ref.~\cite{lumi7Tev}.

The `Modelling' field of Table~\ref{sys_summary} includes the uncertainties
on the methods used for the background estimate, as well
as the modelling uncertainties of the generators used
to assist the estimate.
For \SRZjets\, an additional 20\% uncertainty is assigned to the $ZV$ background estimate to account for the variations 
between data and simulation in the $ZV$ control regions with two or more jets, 
as described in section~\ref{sec:Zjets}.
Uncertainties on the $Z/\gamma^*+\mathrm{jets}$ background estimate in \SRZjets\ 
include 
the systematic uncertainties associated with the jet smearing method due to the fluctuations in
the non-Gaussian tails of the response function and 
the systematic uncertainty associated with the cut value on 
$\MET/\sqrt{\ET^\mathrm{sum}}$ used to define the seed region. 
The effect of using each seed event multiple times is also taken into account.
Generator modelling uncertainties are estimated by comparing the results 
from \powheg\ and \mcatnlo\ generators for top events, and  
\powheg\ and \amcnlo\ for $WW$ events, using \herwig\ for parton showering in all cases.
Parton showering uncertainties are estimated in top and $WW$ events by comparing \powheg\ plus \herwig\ with \powheg\ plus \pythia.
Both generator modelling and parton showering uncertainties are estimated in $ZV$ events by comparing \powheg\ plus \pythia\ to \sherpa.
Special \ttbar\ samples are generated using \acer\ with \pythia\ to evaluate the 
uncertainties related to the amount of initial and final-state radiation~\cite{topjetveto}. 
Impact of the choice of renormalization and factorization scales is evaluated by varying them
between 0.5 and 2 times the nominal values in \powheg\ for top events and \amcnlo\ for diboson events.
The uncertainties due to the PDFs for the top and diboson events are evaluated using 90\% C.L. CT10 PDF eigenvectors.
Effects of using different PDF sets have been found to be negligible.
The dominant contribution among the `Modelling' uncertainties comes from the difference between \powheg\ and  \amcnlo\ for diboson production.

Signal cross-sections are calculated to NLO in the
strong coupling constant. Their uncertainties are taken from an envelope of cross-section 
predictions using different PDF sets and factorization and
renormalization scales, as described in ref.~\cite{Kramer:2012bx}. 
Systematic uncertainties associated with the signal selection efficiency include
those due to lepton trigger, reconstruction and identification, jet reconstruction and
\MET\ calculation.
Uncertainties on the integrated luminosity affect the predicted signal yield.
The total uncertainty on the predicted signal yield is typically 9--13\%
for SUSY scenarios to which this measurement is sensitive.

\section{Results}\label{sec:results}

\begin{figure}
\centering
\raisebox{0.38\textwidth}{(a)}\plot{0.45}{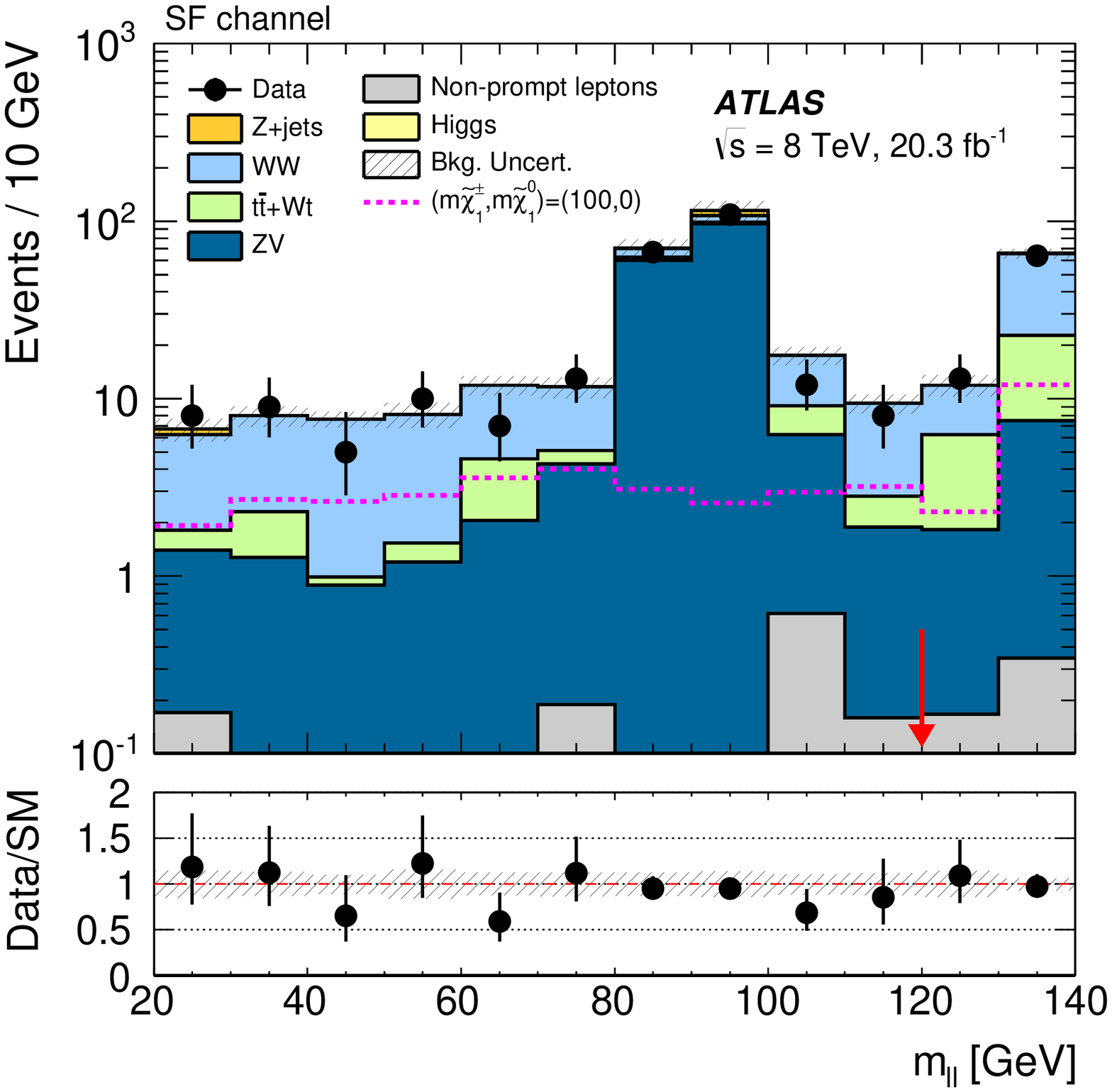}{75,85.5}\quad
\raisebox{0.38\textwidth}{(b)}\plot{0.45}{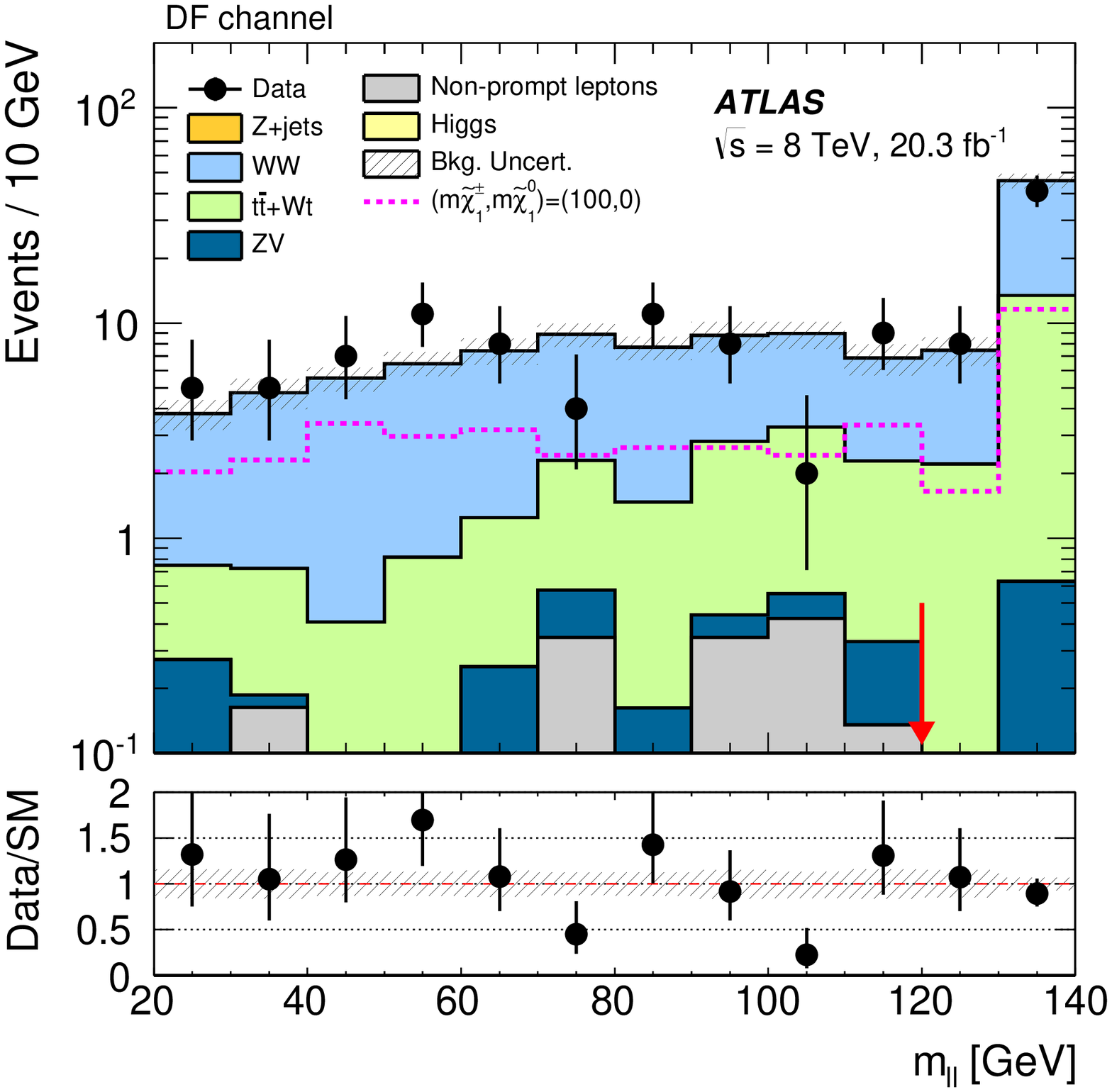}{75,85.5}\\
\raisebox{0.38\textwidth}{(c)}\plot{0.45}{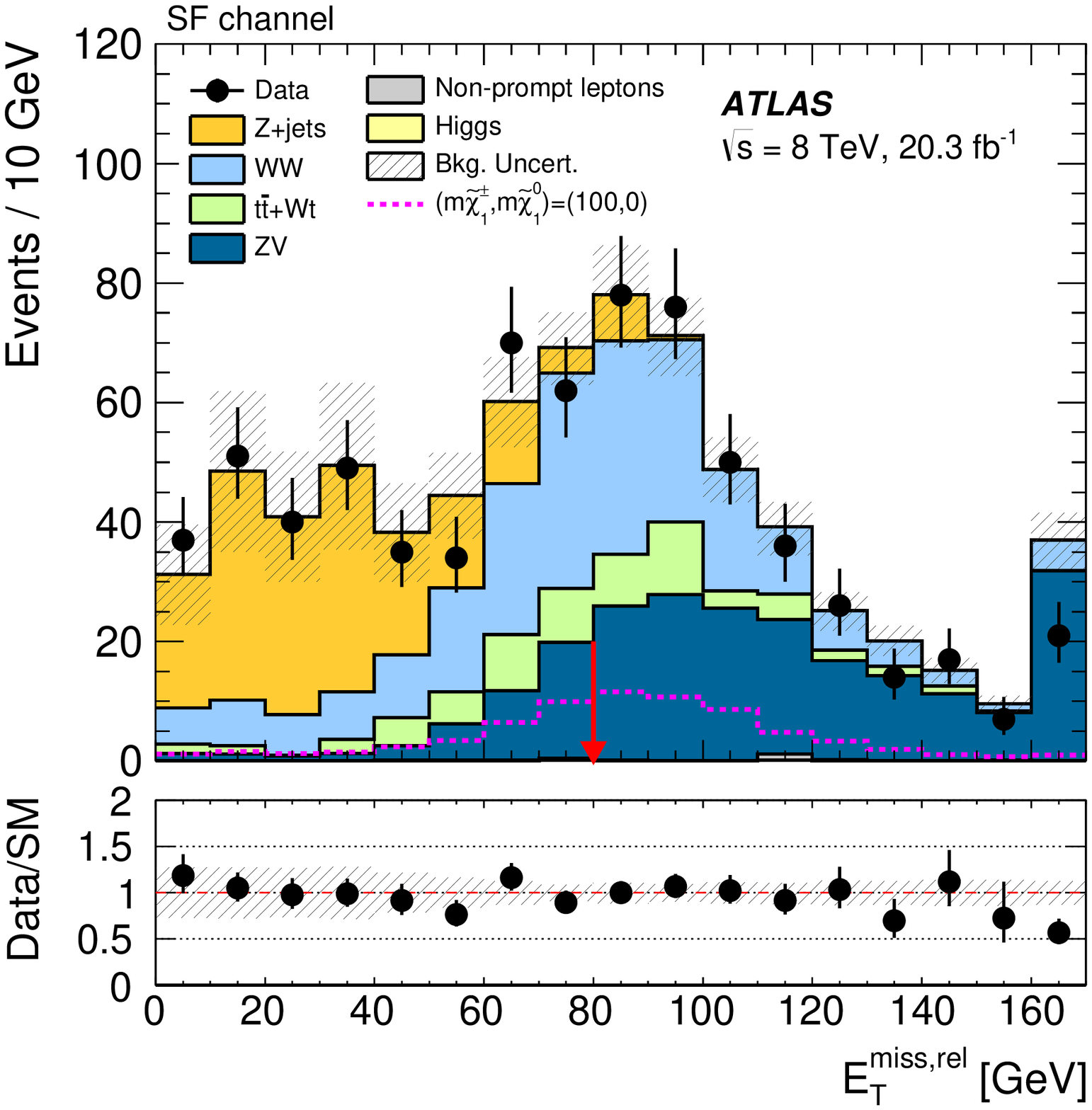}{75,85.5}\quad
\raisebox{0.38\textwidth}{(d)}\plot{0.45}{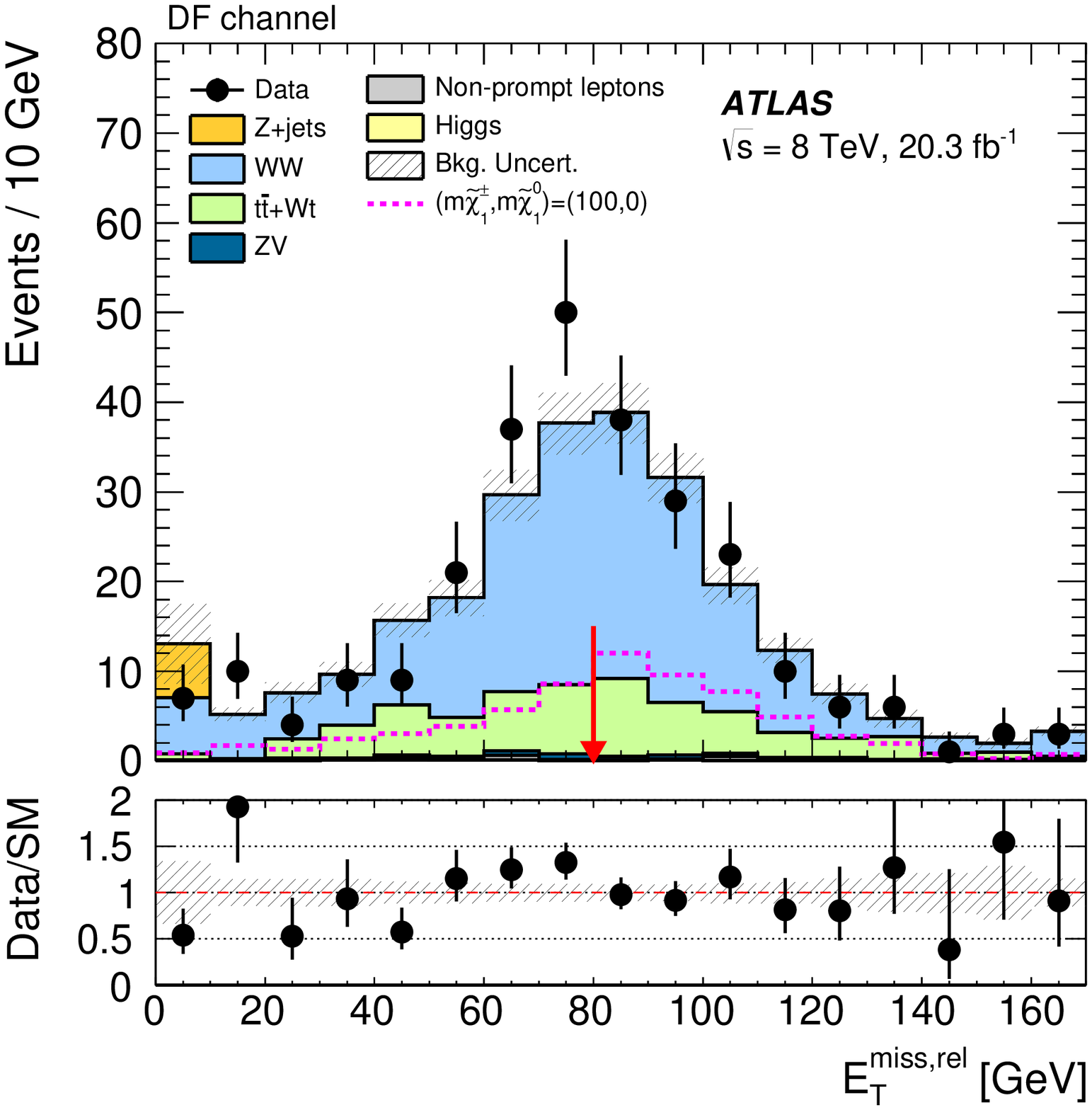}{75,85.5}
\caption{\label{fig:SRdist1}
	Distributions of \mll\ in the (a) SF and (b) DF samples that satisfy all the
	\SRWWa\ selection criteria except for the one on \mll,
	and of \METrel\ in the (c) SF and (d) DF samples that satisfy all the
	\SRWWa\ selection criteria except for the ones on \mll\ and \METrel.
The lower panel of each plot shows the ratio between data and the SM background
prediction.
The hashed regions represent the sum in quadrature of systematic uncertainties
and statistical uncertainties arising from the numbers of MC events. 
	Predicted signal distributions in a simplified model with $\mCone=100\GeV$ and $\mNone=0$ 
	are superimposed.
	Red arrows indicate the \SRWWa\ selection criteria.
	In (a), the region $81.2<\mll<101.2\GeV$ is rejected by the $Z$ boson veto.
}
\end{figure}

\begin{figure}
\centering
\raisebox{0.38\textwidth}{(a)}\plot{0.45}{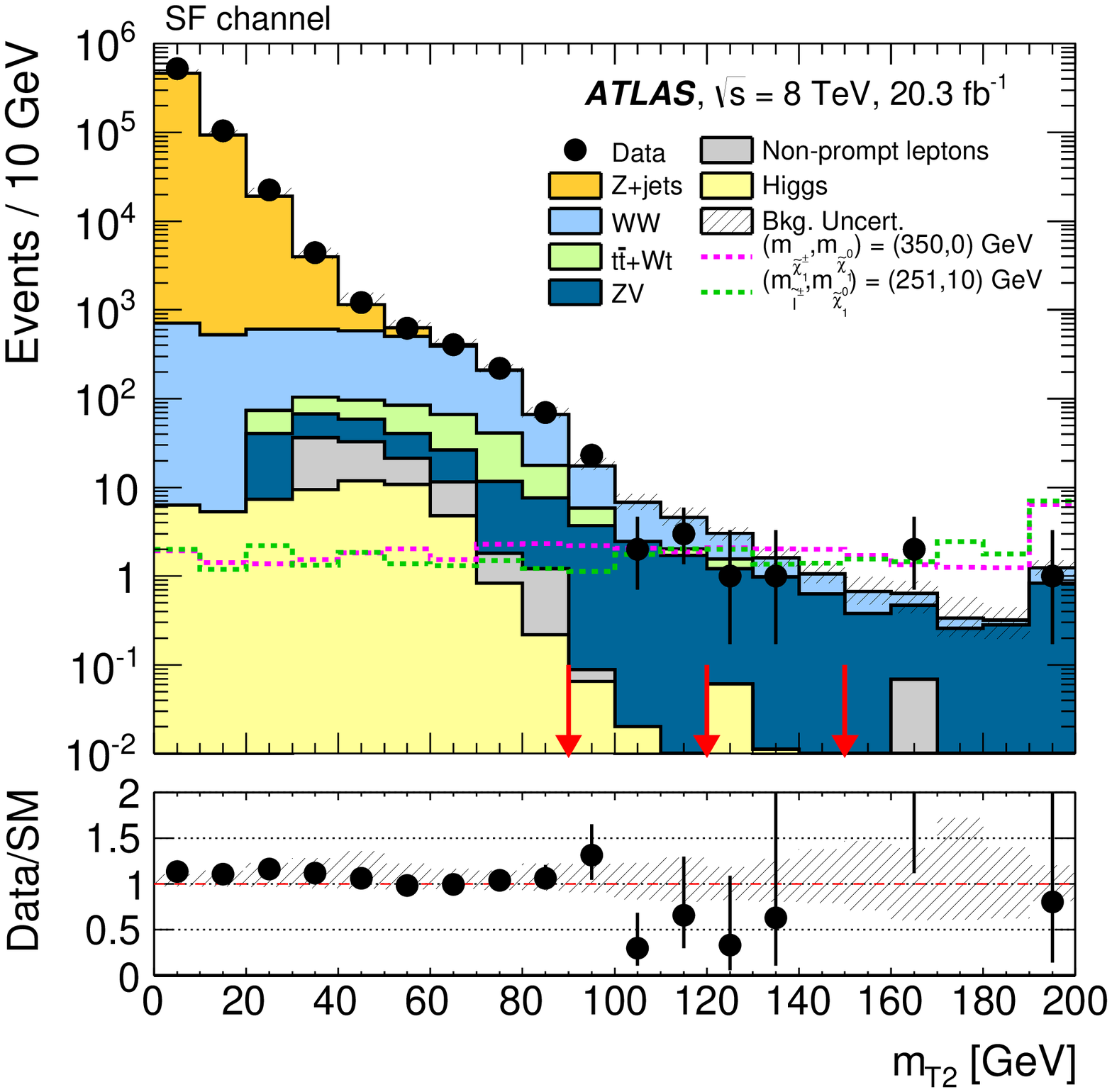}{40,86}
\raisebox{0.38\textwidth}{(b)}\plot{0.45}{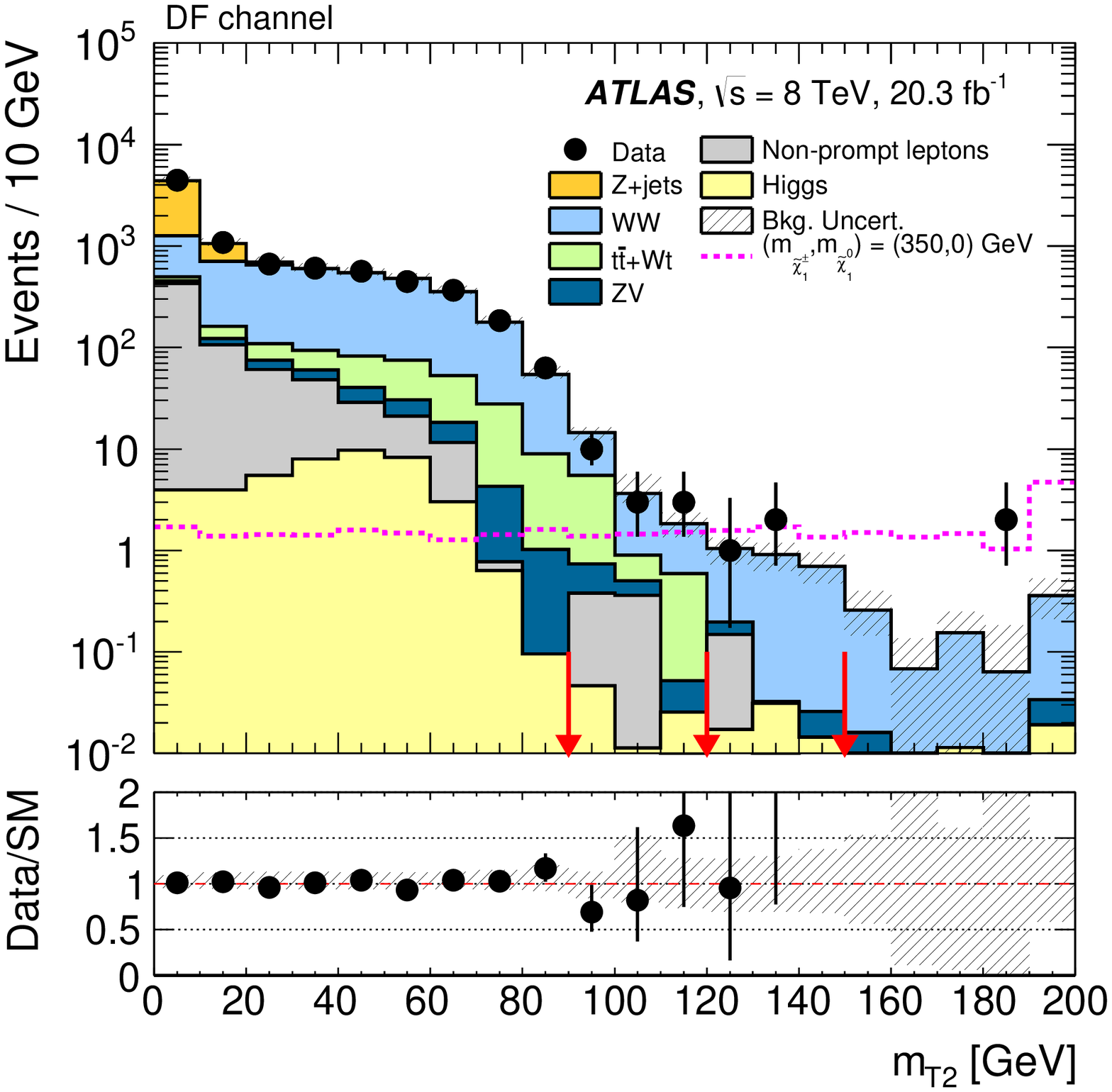}{40,86}
\raisebox{0.38\textwidth}{(c)}\plot{0.45}{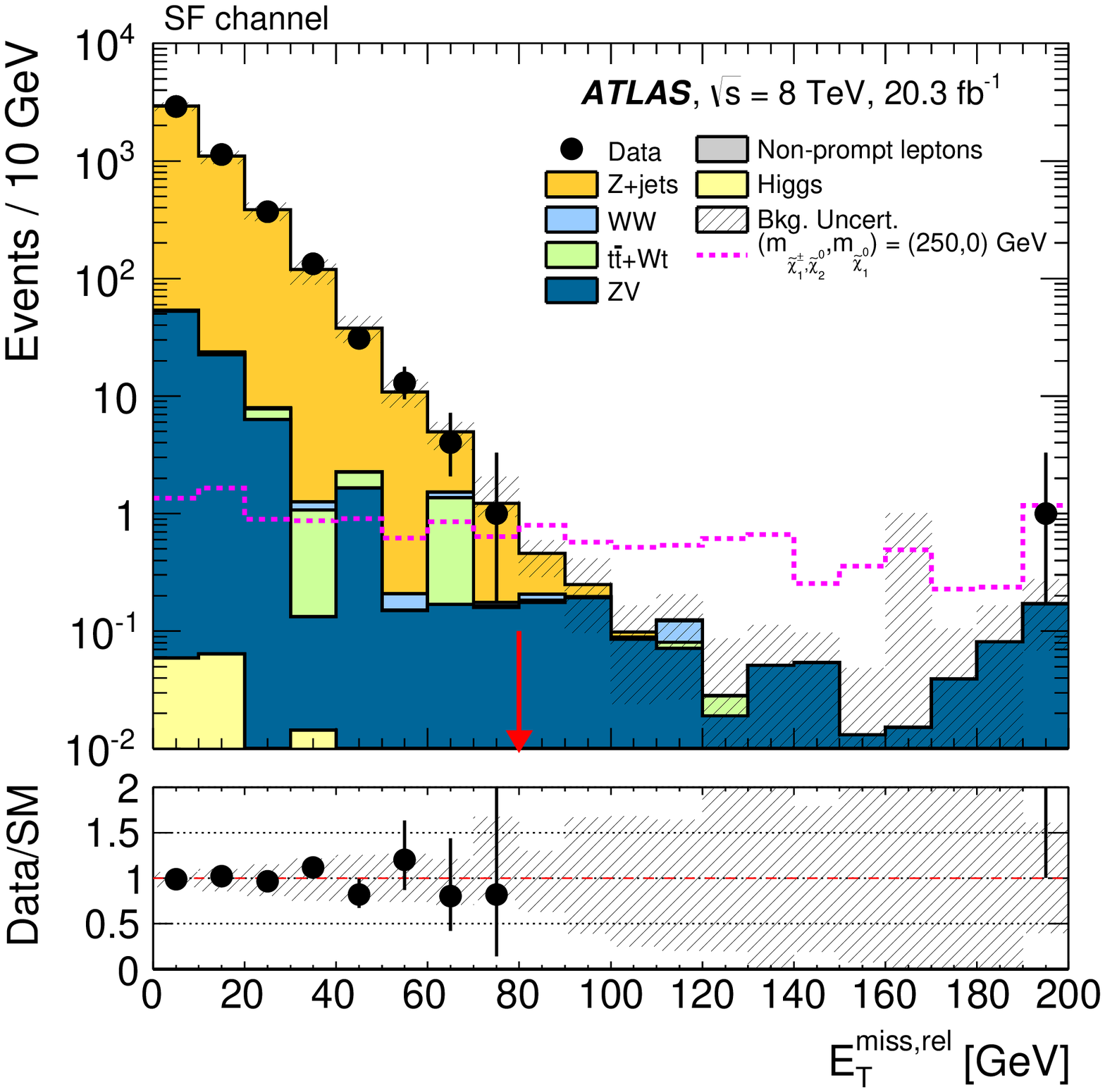}{40,86}
\caption{\label{fig:SRdist2}
	Distributions of \mttwo\ in the (a) SF and (b) DF samples that satisfy all
	the \SRmttwo\ selection criteria except for the one on \mttwo,
	and of (c) \METrel\ in the sample that satisfies all the \SRZjets\ selection criteria
	except for the one on \METrel.
The lower panel of each plot shows the ratio between data and the SM background
prediction.
The hashed regions represent the sum in quadrature of systematic uncertainties
and statistical uncertainties arising from the numbers of MC events. 
	Predicted signal distributions in simplified models with 
	$\mCone=350\GeV$, $\mslepton=\msneutrino=175\GeV$ and $\mNone=0$ are
	superimposed in (a) and (b),
	$\mslepton=251\GeV$ and $\mNone=10\GeV$ in (a),
	and $\mCone=\mNtwo=250\GeV$ and $\mNone=0$ in (c).
	Red arrows indicate the selection criteria
	for \SRmttwo\ and \SRZjets.
}
\end{figure}

Figures~\ref{fig:SRdist1} and \ref{fig:SRdist2}
show the comparison between data and the SM prediction for key kinematic variables 
in different signal regions.
In each plot, the expected distributions from the $WW$, \ttbar\ and $ZV$ processes are
corrected with data-driven normalization factors obtained from the fit detailed in section~\ref{sec:backgrounds}.
The hashed regions represent the sum in quadrature of systematic uncertainties
 and statistical uncertainties 
arising from the numbers of MC events. 
The effect of limited data events in the CR is included in the systematic uncertainty. 
All statistical uncertainties are added in quadrature whereas the systematic 
uncertainties are obtained after taking full account of all correlations between sources, 
background contributions and channels. 
The rightmost bin of each plot includes overflow.
Illustrative SUSY benchmark models, normalized to the integrated luminosity, are superimposed.
The lower panel of each plot shows the ratio between data and the SM background
prediction.

Tables~\ref{tab:expmt2}, \ref{tab:expww} and \ref{tab:expzjets} compare the 
observed yields in each signal region with those predicted for the SM background.
The errors include both statistical and systematic uncertainties.
Good agreement is observed across all channels.

For each SR, 
the significance of a possible excess over the SM background 
is quantified by the one-sided probability, $p_0$,
of the background alone to fluctuate to the observed number of events or higher,
using the asymptotic formula~\cite{statforumlimitsl}.
This is calculated using a fit similar to the one described in section~\ref{sec:fit},
but including the observed number of events in the SR as an input.
The accuracy of the limits obtained from the asymptotic formula was tested for all SRs by randomly generating a large number of pseudo data sets and repeating the fit.
Upper limits at 95\% CL on the number of non-SM events for each SR are 
derived using the \CLs\ prescription~\cite{Read:2002hq} 
and neglecting any possible contamination in the CRs. 
Normalizing these by the integrated luminosity of the data sample 
they can be interpreted as upper limits, \limsigvis, on the visible non-SM cross-section, 
defined as the product 
of acceptance, reconstruction efficiency and production cross-section of the non-SM contribution.
All systematic uncertainties and their correlations are taken into account via nuisance parameters.
The results are given in 
Tables~\ref{tab:expmt2}, \ref{tab:expww} and \ref{tab:expzjets}.

\begin{table}\small
\begin{center} 
\caption{\label{tab:expmt2}
Observed and expected numbers of events in \SRmttwo.
Also shown are the one-sided $p_0$ values 
and the observed and expected 95\% CL upper limits, \limsigvis,
on the visible cross-section for non-SM events.
The `Others' background category includes \fake\ lepton, $Z/\gamma^*+\mathrm{jets}$ and SM Higgs.  
The numbers of signal events are shown for the $\gaugino{1}{+}\gaugino{1}{-}\to(\slepton\nu\text{ or }\ell\sneutrino)\neutralino{1}(\slepton'\nu'\text{ or }\ell'\sneutrino')\neutralino{1}$ scenario and for the
$\slepton^+\slepton^-\to\ell^+\neutralino{1}\ell^-\neutralino{1}$ scenario with different \chargino{1}, \neutralino{1} and \slepton\ masses in GeV.
}\smallskip
\begin{tabular}{l|CC|CC|CC}
\hline
& \multicolumn{2}{c|}{\SRmtt{90}} & \multicolumn{2}{c|}{\SRmtt{120}} & \multicolumn{2}{c}{\SRmtt{150}} \\
& $SF$ & $DF$ & $SF$ & $DF$ & $SF$ & $DF$ \\
\hline
Expected background &&&&&& \\
\quad$WW$         & 22.1\pm4.3 & 16.2\pm3.2 & 3.5\pm1.3       & 3.3\pm1.2 & 1.0\pm0.5 & 0.9\pm0.5 \\
\quad$ZV$         & 12.9\pm2.2 & \spz0.8\pm0.2 & 4.9\pm1.6       & 0.2\pm0.1 & 2.2\pm0.5 & <0.1      \\
\quad Top         & \spz3.0\pm1.8 & \spz5.5\pm1.9 & 0.3^{+0.4}_{-0.3} & <0.1    & <0.1      & <0.1      \\
\quad Others      & \spz0.3\pm0.3 & \spz0.8\pm0.6 & 0.1^{+0.4}_{-0.1} & 0.1\pm0.1 & 0.1^{+0.4}_{-0.1} & 0.0^{+0.4}_{-0.0} \\
Total             & 38.2\pm5.1 & 23.3\pm3.7 & 8.9\pm2.1       & 3.6\pm1.2 & 3.2\pm0.7 & 1.0\pm0.5 \\
\hline
Observed events   & 33         & 21         & 5               & 5         & 3         & 2 \\
\hline
Predicted signal &&&&&& \\
$(m_{\chargino{1}},m_{\neutralino{1}})=(350,0)$
               & 24.2\pm2.5 & 19.1\pm2.1 & 18.1\pm1.8 & 14.7\pm1.7 & 12.0\pm1.3 & 10.1\pm1.3 \\
$(m_{\slepton},m_{\neutralino{1}})=(251,10)$
               & 24.0\pm2.7 & $---$ & 19.1\pm2.5 & $---$ & 14.3\pm1.7 & $---$ \\
\hline
$p_0$ & 0.50 & 0.50 & 0.50 & 0.27 & 0.50 & 0.21 \\
Observed \limsigvis\ [fb] & 0.63 & 0.55 & 0.26 & 0.36 & 0.24 & 0.26 \\
Expected \limsigvis\ [fb] & 0.78^{+0.32}_{-0.23} & 0.62^{+0.26}_{-0.18} 
                          & 0.37^{+0.17}_{-0.11} & 0.30^{+0.13}_{-0.09} 
                          & 0.24^{+0.13}_{-0.08} & 0.19^{+0.10}_{-0.06} \\
\hline
\end{tabular}
\end{center}
\end{table}

\begin{table}\small\centering
\caption{\label{tab:expww}
Observed and expected numbers of events in \SRWW.
Also shown are the one-sided $p_0$ values 
and the observed and expected 95\% CL upper limits, \limsigvis,
on the visible cross-section for non-SM events.
The `Others' category includes \fake\ lepton, $Z/\gamma^*+\mathrm{jets}$ and SM Higgs.  
The numbers of signal events are shown for the $\gaugino{1}{+}\gaugino{1}{-}\to W^+\neutralino{1}W^-\neutralino{1}$ scenario with different \chargino{1} and \neutralino{1} masses in GeV\@.
}\smallskip
\begin{tabular}{l|CC|CC|CC}
\hline
& \multicolumn{2}{c|}{\SRWWa} & \multicolumn{2}{c|}{\SRWWb} & \multicolumn{2}{c}{\SRWWc} \\
& $SF$ & $DF$ & $SF$ & $DF$ & $SF$ & $DF$ \\
\hline
Background &&&&&& \\
\quad$WW$         & 57.8\pm5.5 & 58.2\pm6.0 & 16.4\pm2.5 & 12.3\pm2.0 & 10.4\pm2.7       & 7.3\pm1.9 \\
\quad$ZV$         & 16.3\pm3.5 & \spz1.8\pm0.5 & 10.9\pm1.9 & \spz0.6\pm0.2 & \spz9.2\pm2.1       & 0.4\pm0.2 \\     
\quad Top         & \spz9.2\pm3.5 & 11.6\pm4.3 & \spz2.4\pm1.7 & \spz4.3\pm1.6 &  0.6^{+1.2}_{-0.6} & 0.9\pm0.8 \\
\quad Others      & \spz3.3\pm1.5 & \spz2.0\pm1.1 & \spz0.5\pm0.4 & \spz0.9\pm0.6 &  0.1^{+0.5}_{-0.1}       & 0.4\pm0.3 \\
Total             & 86.5\pm7.4 & 73.6\pm7.9 & 30.2\pm3.5 & 18.1\pm2.6 & 20.3\pm3.5       & 9.0\pm2.2 \\
\hline
Observed events & 73 & 70 & 26 & 17 & 10 & 11 \\
\hline
Predicted signal &&&&&&\\
$(m_{\chargino{1}},m_{\neutralino{1}})=(100,0)$
               & 25.6\pm3.3 & 24.4\pm2.2   &  &  &  & \\
$(m_{\chargino{1}},m_{\neutralino{1}})=(140,20)$
               & & & 8.3\pm0.8          & 7.2\pm0.8              & & \\
$(m_{\chargino{1}},m_{\neutralino{1}})=(200,0)$
               & & & & & 5.2\pm0.5          & 4.6\pm0.4              \\
\hline
$p_0$ & 0.50 & 0.50 & 0.50 & 0.50 & 0.50 & 0.31 \\
Observed \limsigvis\ [fb] & 0.78 & 1.00 & 0.54 & 0.49 & 0.29 & 0.50 \\
Expected \limsigvis\ [fb] & 1.13^{+0.46}_{-0.32} & 1.11^{+0.44}_{-0.31} 
                          & 0.66^{+0.28}_{-0.20} & 0.53^{+0.23}_{-0.16} 
                          & 0.52^{+0.23}_{-0.16} & 0.41^{+0.19}_{-0.12} \\
\hline
\end{tabular}
\end{table}

\begin{table}\small\centering
\caption{\label{tab:expzjets}
Observed and expected numbers of events in \SRZjets.
Also shown are the one-sided $p_0$ value and the
observed and expected 95\% CL upper limits, \limsigvis,
on the visible cross-section for non-SM events.
The numbers of signal events are shown for the $\chargino{1}\neutralino{2}\to W^{\pm}\neutralino{1}Z\neutralino{1}$ scenario with different \chargino{1}, \neutralino{2} and \neutralino{1} masses in GeV\@.
}\smallskip
\begin{tabular}{l|C}
\hline
& \text{\SRZjets} \\ 
\hline
Background & \\
\quad$WW$         & 0.1\pm0.1 \\
\quad$ZV$         & 1.0\pm0.6 \\
\quad Top         & <0.1      \\
\quad $Z$ + jets and others & 0.3\pm0.2 \\
Total             & 1.4\pm0.6 \\
\hline
Observed events & 1 \\
\hline
Predicted signal &\\
$(m_{\neutralino{2},\chargino{1}},m_{\neutralino{1}})=(250,0)$ & 6.4\pm0.8 \\
$(m_{\neutralino{2},\chargino{1}},m_{\neutralino{1}})=(350,50)$ & 3.7\pm0.2 \\
\hline
$p_0$ & 0.50 \\
Observed \limsigvis\ [fb] & 0.17 \\
Expected \limsigvis\ [fb] & 0.19^{+0.11}_{-0.06} \\
\hline
\end{tabular}
\end{table}

\section{Interpretation}

Exclusion limits at 95\% confidence-level are set on the slepton, chargino and neutralino masses within the specific scenarios considered. 
The same \CLs\ limit-setting procedure as in section~\ref{sec:results} is used, except that the 
SUSY signal is allowed to populate both the signal region and the control regions as predicted by 
the simulation.
Since the SRs are not mutually exclusive, the SR with the best expected exclusion limit is chosen for each model point.

\begin{figure}\centering
	\plot{0.45}{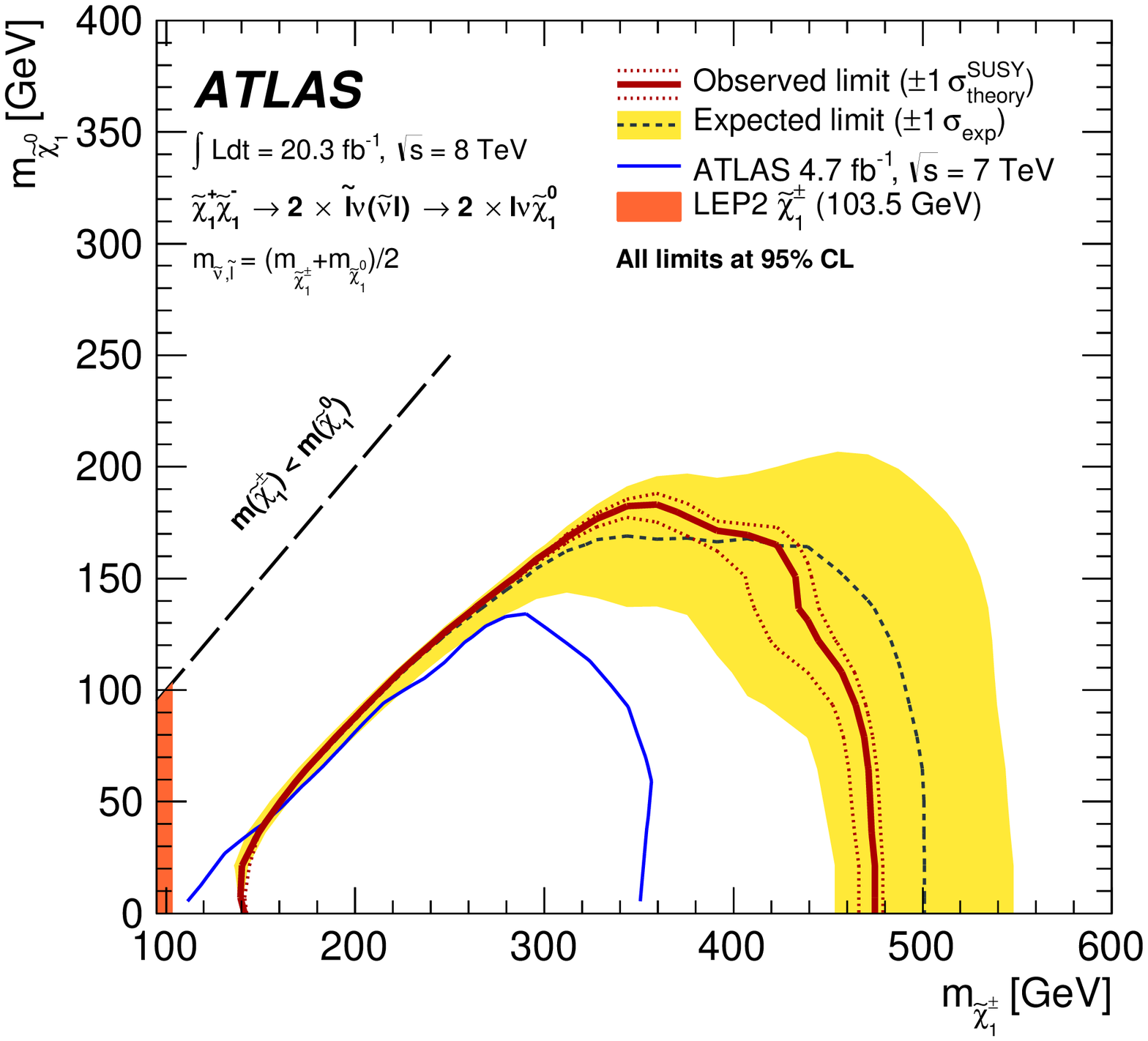}{33,77.5}
	\caption{\label{interpretation_modeA}
		Observed and expected
		95\% CL exclusion regions in the \mCone--\mNone\ plane for simplified-model \ConeCone\ 
		pair production with common masses of sleptons and sneutrinos at
		$m_{\slepton}=m_{\sneutrino}=(\mCone+\mNone)/2$.
		Also shown is the LEP limit~\cite{lepsusy} on the mass of the chargino.
		The blue line indicates the limit from the previous analysis with the 7\,TeV 
		data~\cite{Aad:2012pxa}.
	}
\end{figure}

\begin{figure}\centering
	\raisebox{0.36\textwidth}{(a)} \plot{0.45}{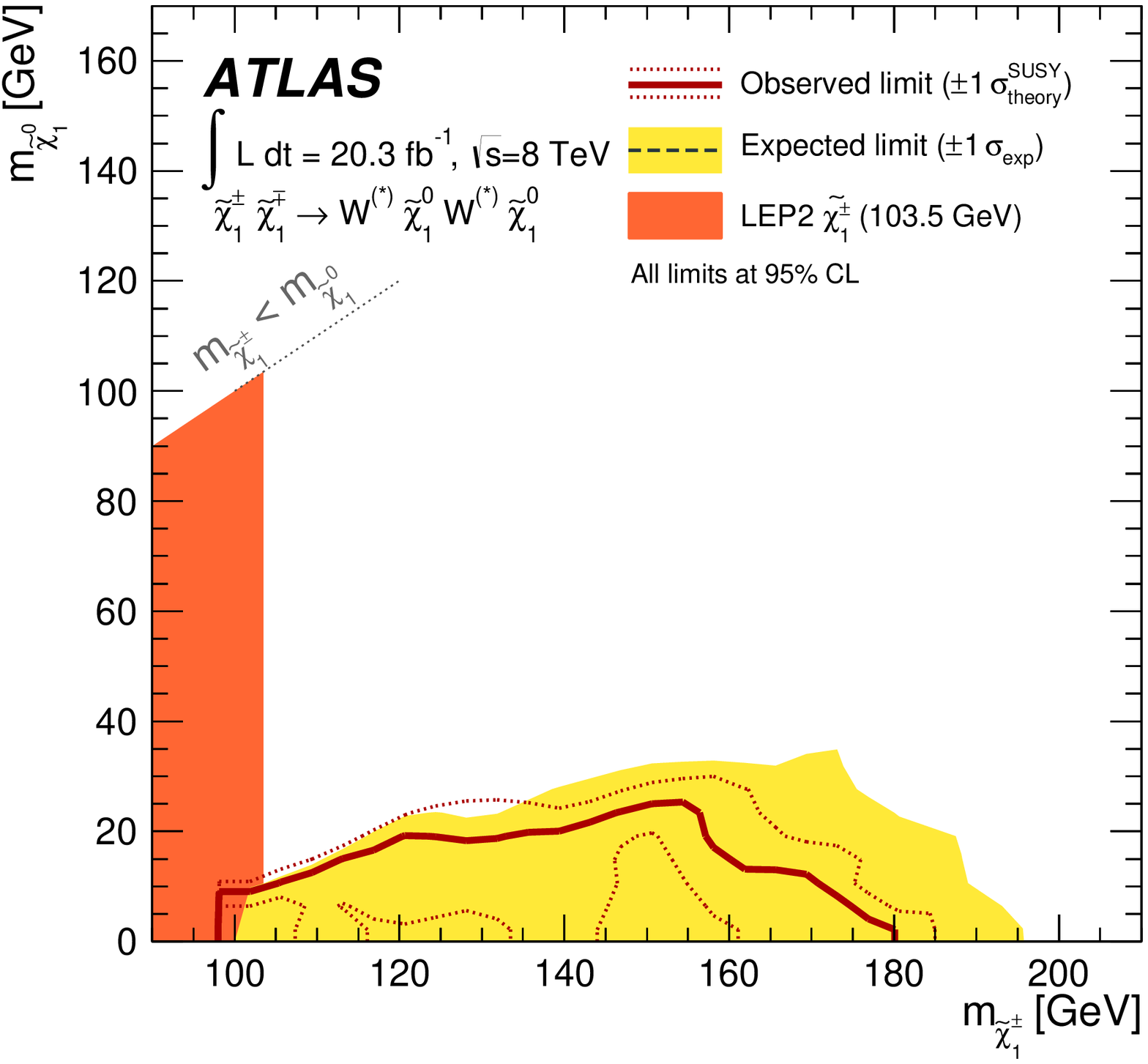}{33,77.5}
	\raisebox{0.36\textwidth}{(b)} \plot{0.45}{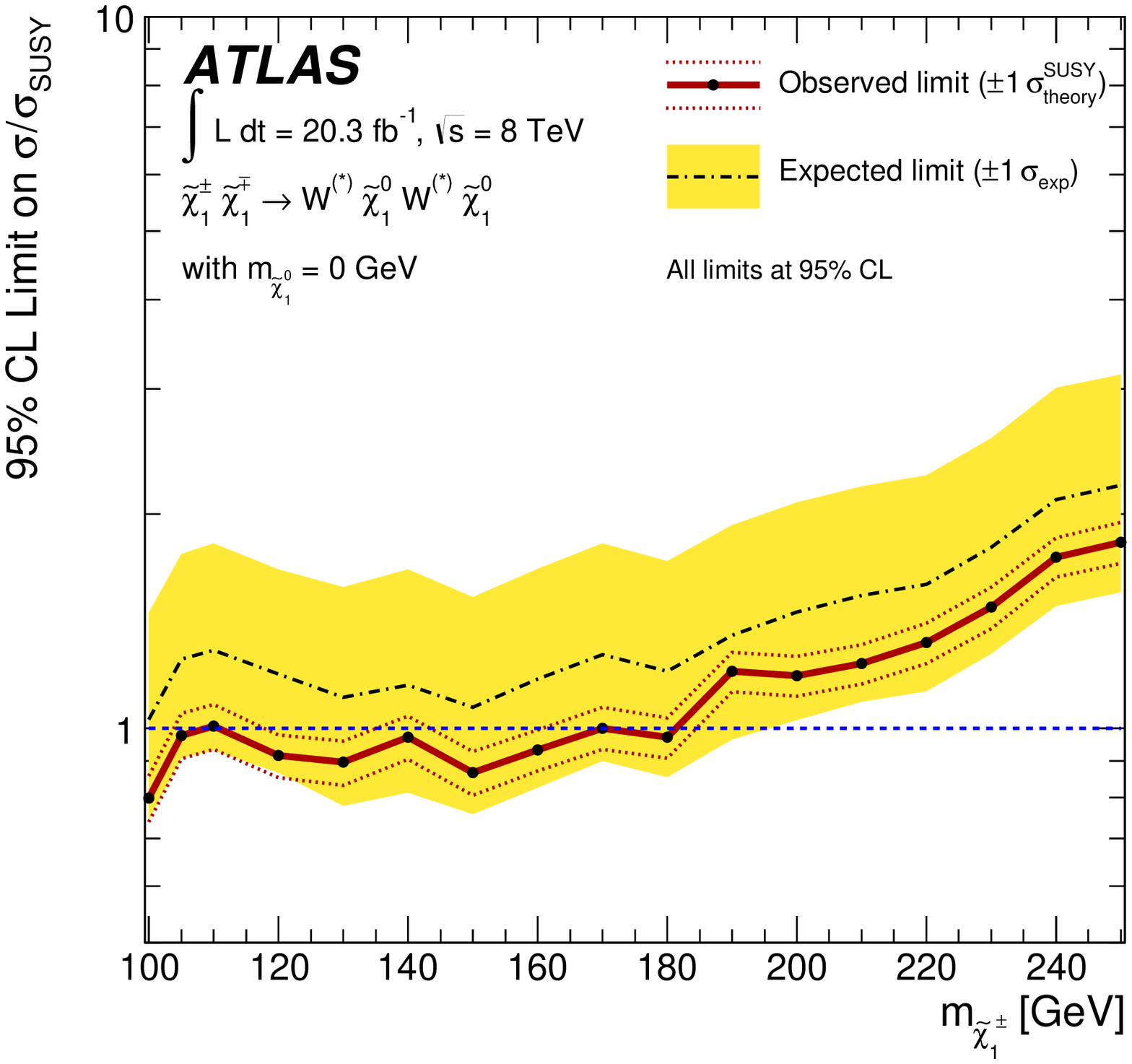}{33,79.5}
	\caption{\label{fig:WWFinal}
		 (a) Observed and expected 95\% CL exclusion regions in the \mNone--\mCone\ plane for simplified-model 
		 \ConeCone\ production followed by $W$-mediated decays.
		 		Also shown is the LEP limit~\cite{lepsusy} on the mass of the chargino.
		(b) Observed and expected 95\% CL upper limits on the cross-section normalized
		by the simplified-model prediction as a function of \mCone\ for $\mNone=0$.}
\end{figure}

\begin{figure}\centering
	\raisebox{0.36\textwidth}{(a)} \plot{0.45}{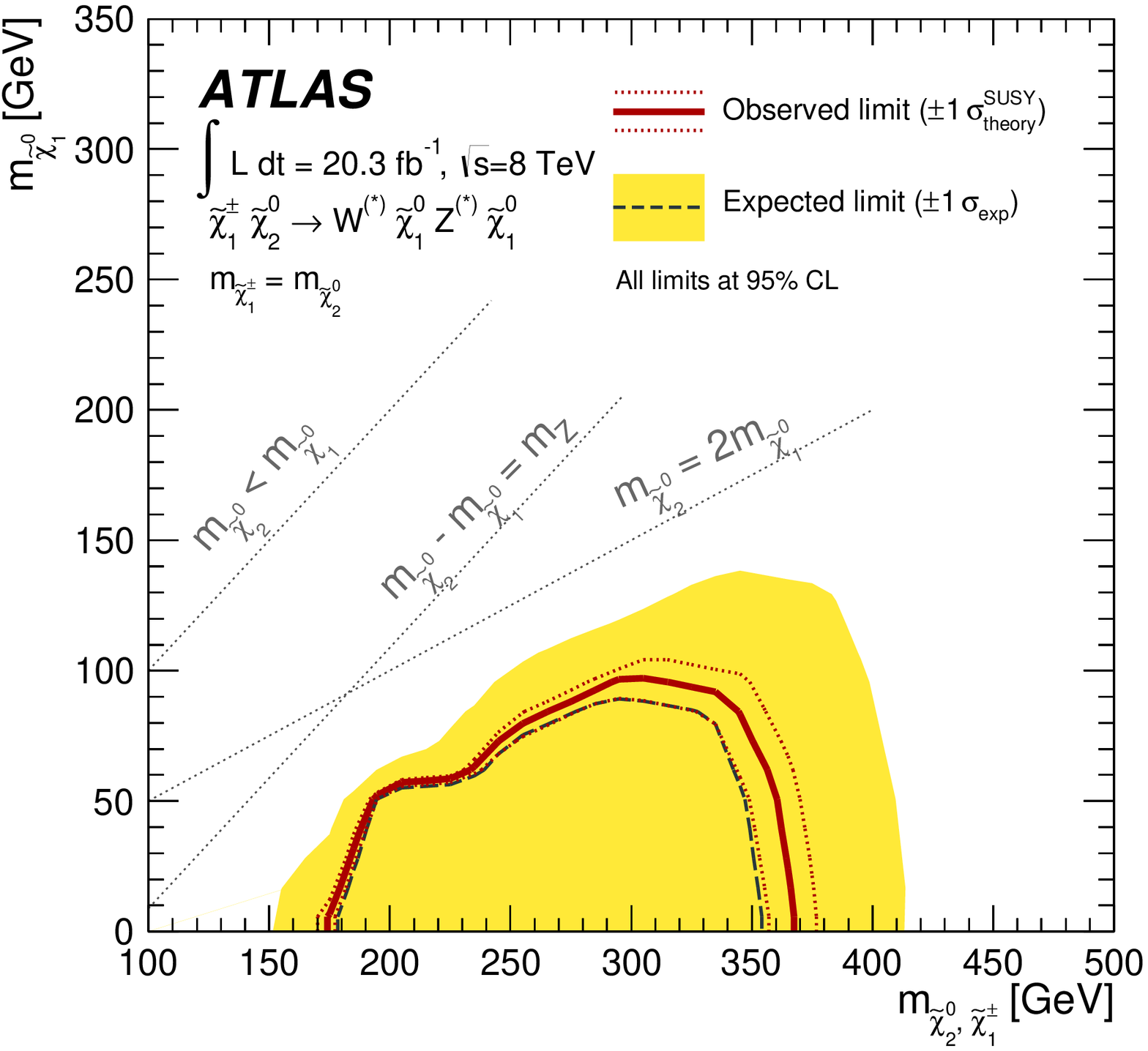}{33,77.5}
	\raisebox{0.36\textwidth}{(b)} \plot{0.45}{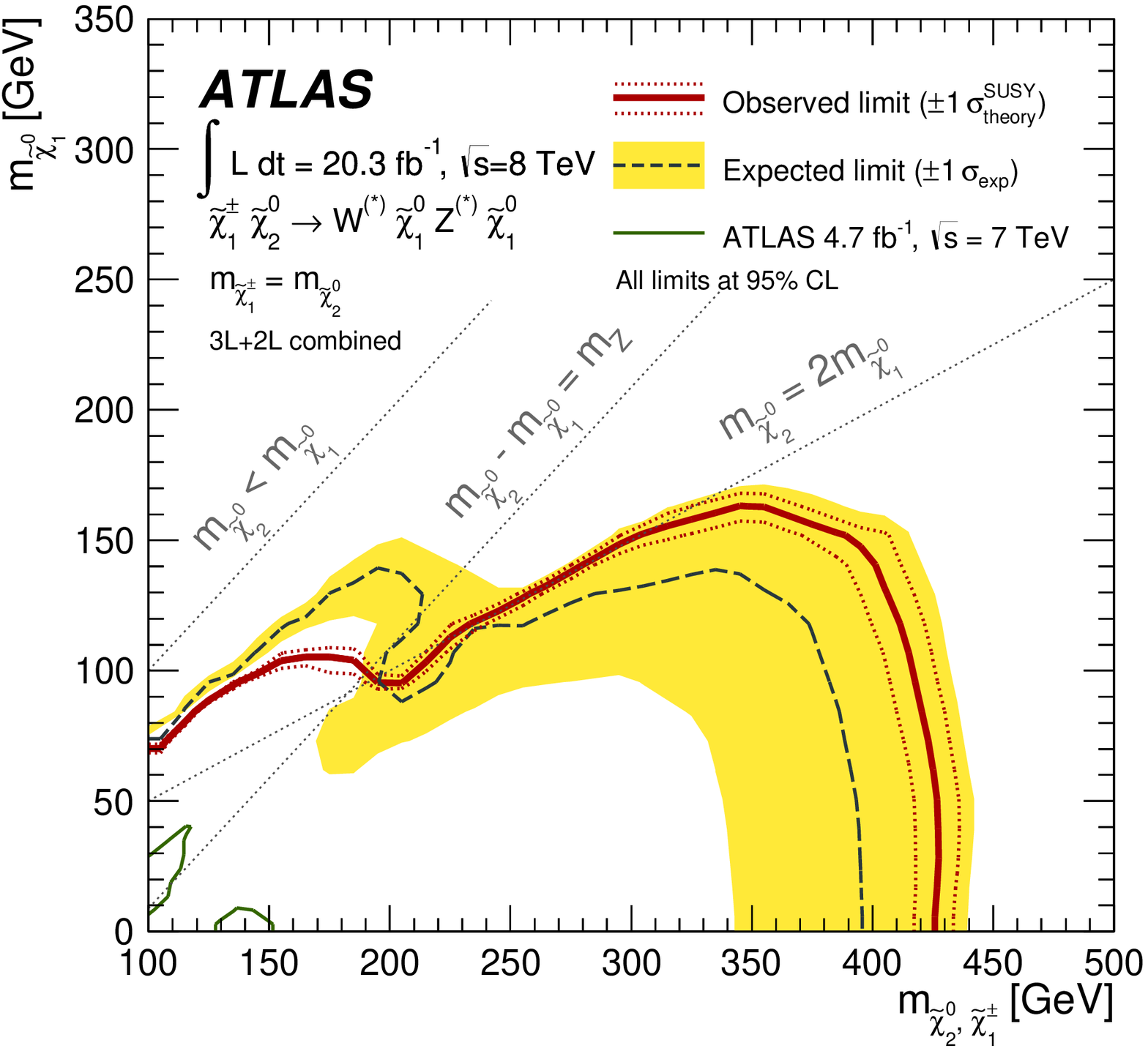}{33,77.5}
	\caption{\label{interpretation_modeC}
		(a) Observed and expected 95\% CL exclusion regions in the $m_{\smash{\Ntwo,\Cone}}$--\mNone\ plane
		for simplified-model \chargino{1}\neutralino{2} production
		followed by $W$ and $Z$-mediated decays obtained from \SRZjets; and 
		(b) the exclusion regions obtained by combining with the ATLAS three-lepton search~\cite{ATLAS_3LEWK2013}.
		The green lines in (b) indicate the regions excluded by ATLAS
	    using $4.7\,\ifb$ of $\sqrt{s}=7\TeV$ data~\cite{ATLAS_3L7TeV}.
	}
\end{figure}

\begin{figure}\centering
	\raisebox{0.36\textwidth}{(a)} \plot{0.45}{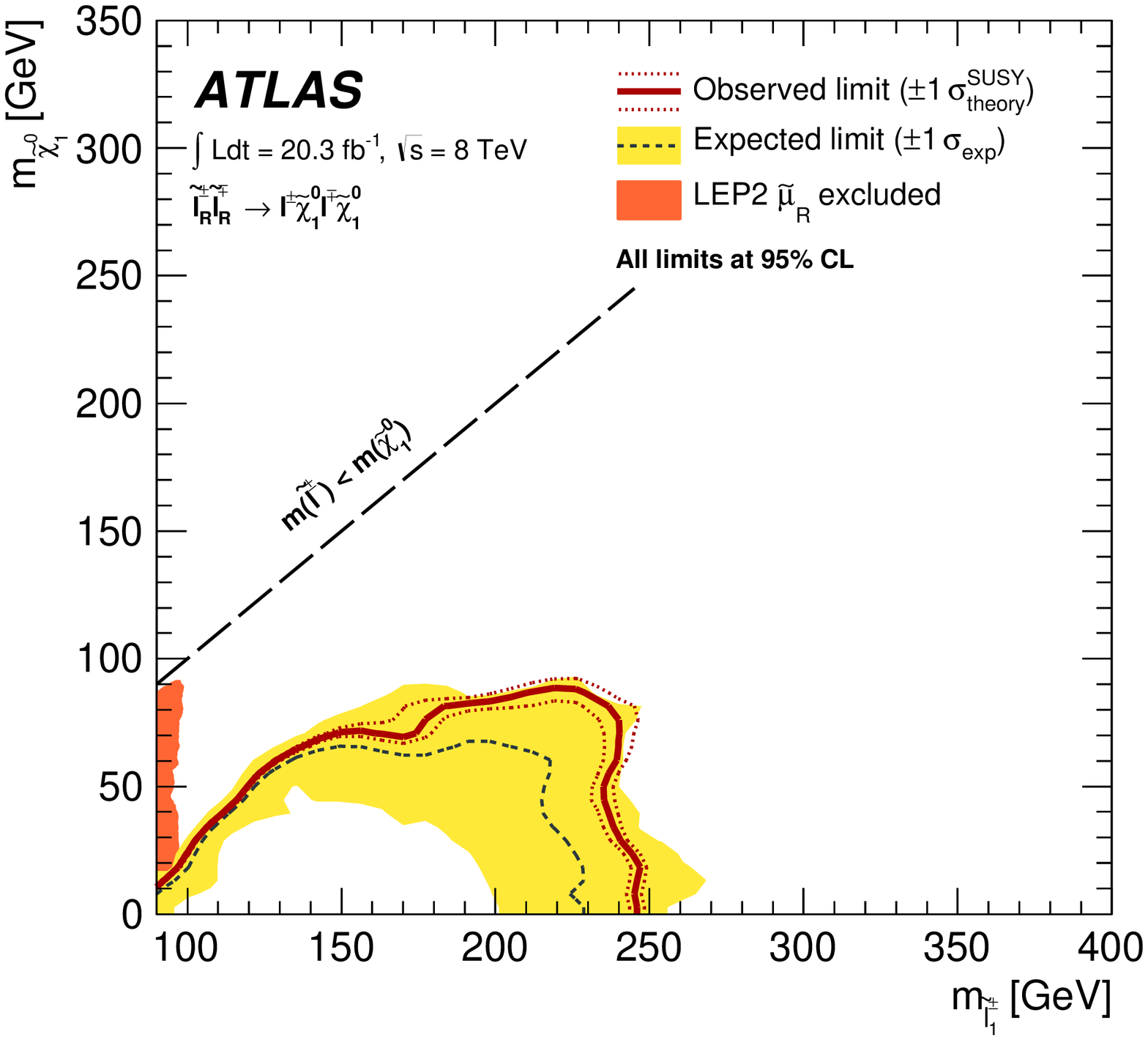}{33,77.5}
	\raisebox{0.36\textwidth}{(b)} \plot{0.45}{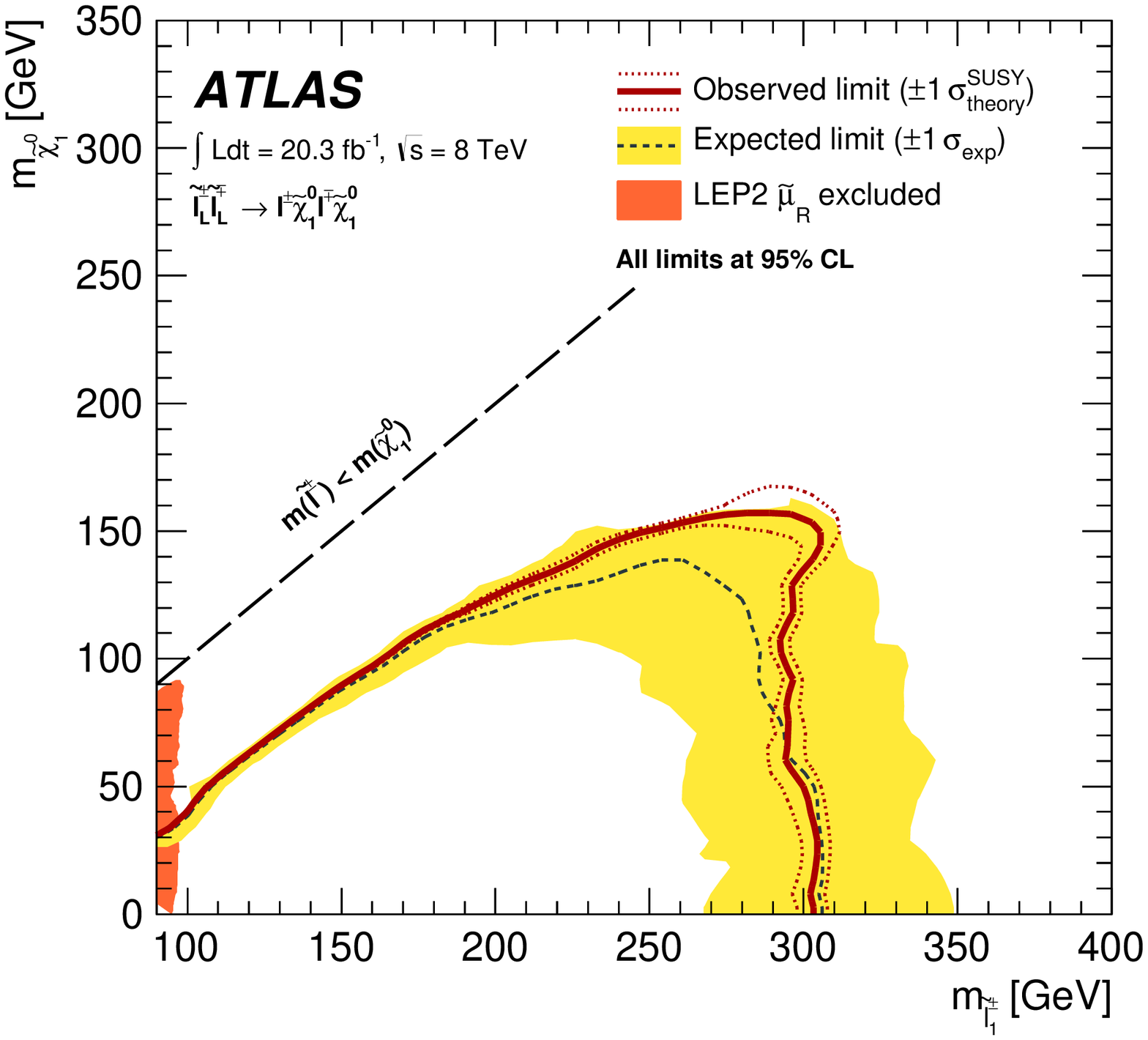}{33,77.5}
	\raisebox{0.36\textwidth}{(c)} \plot{0.45}{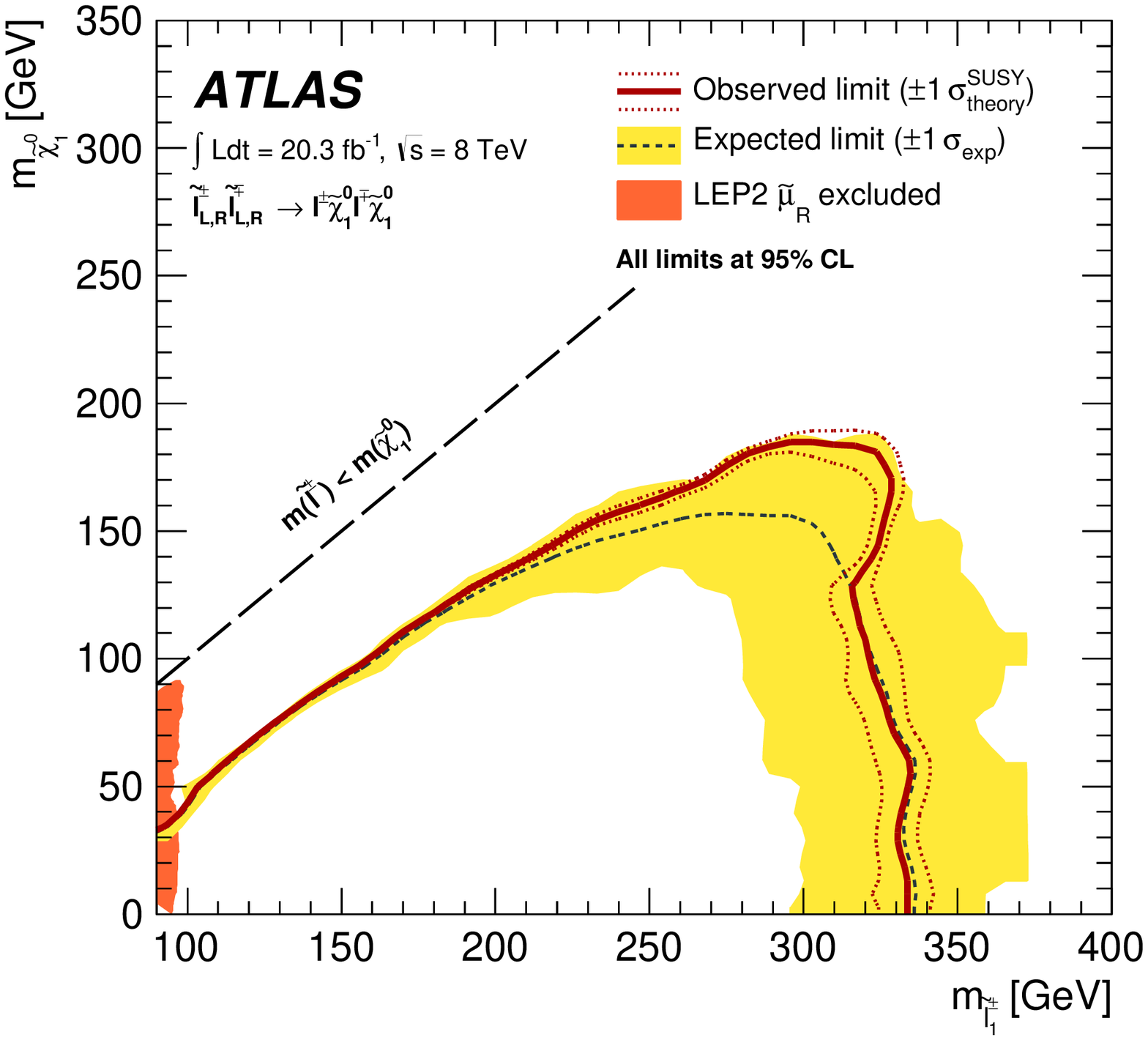}{33,77.5}
	\caption{\label{interpretation_slepton}
		95\% CL exclusion regions in the \mNone--$m_{\slepton}$ plane for 
		(a) right-handed, (b) left-handed, and (c) both right- and left-handed (mass degenerate)
		selectron and smuon production. 
		Also illustrated are the LEP limits~\cite{lepsusy} on the mass of the right-handed 
		smuon $\tilde{\mu}_{R}$.
		}
\end{figure}

\begin{figure}\centering
	\raisebox{0.36\textwidth}{(a)} \plot{0.45}{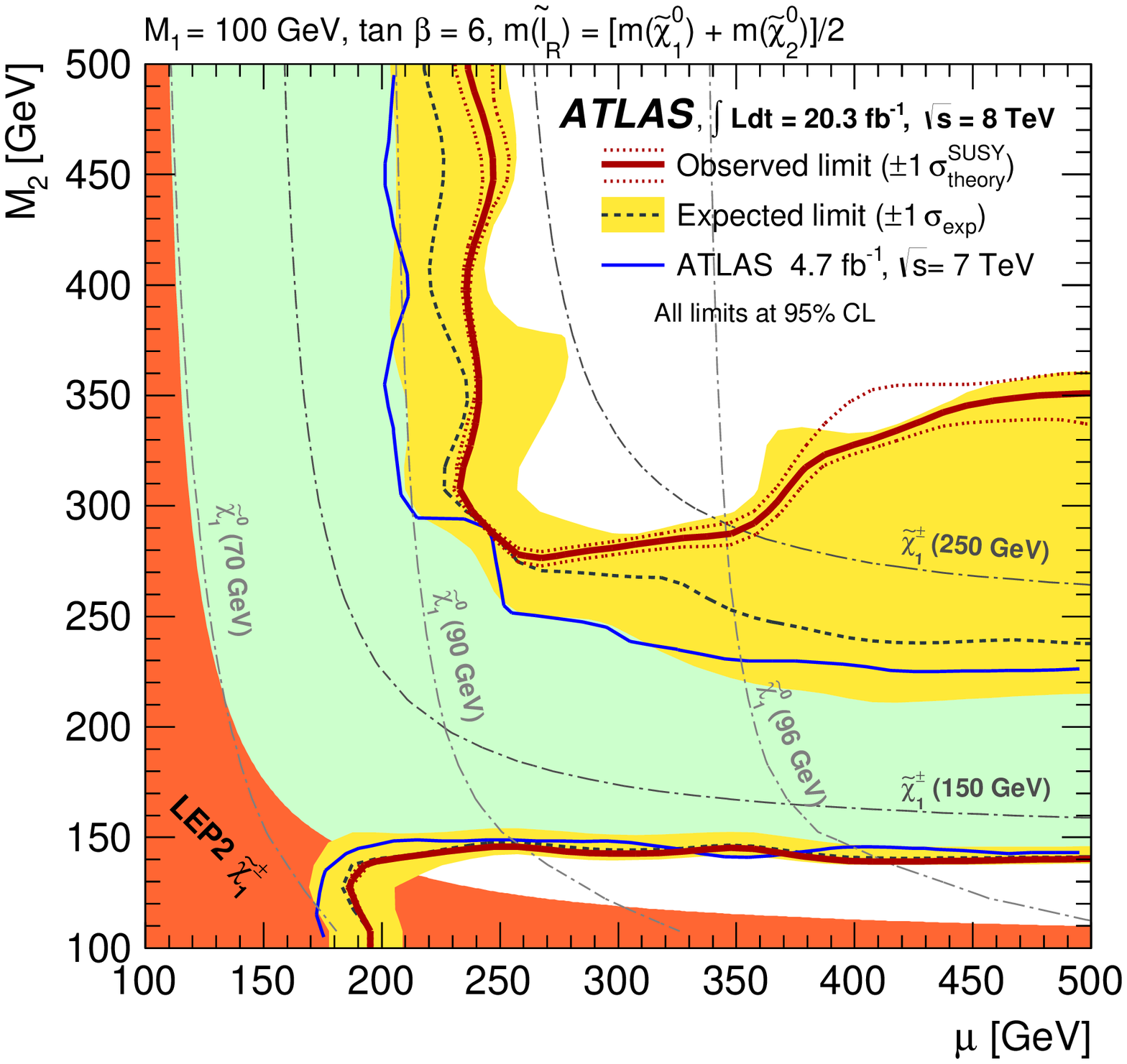}{72,79}
	\raisebox{0.36\textwidth}{(b)} \plot{0.45}{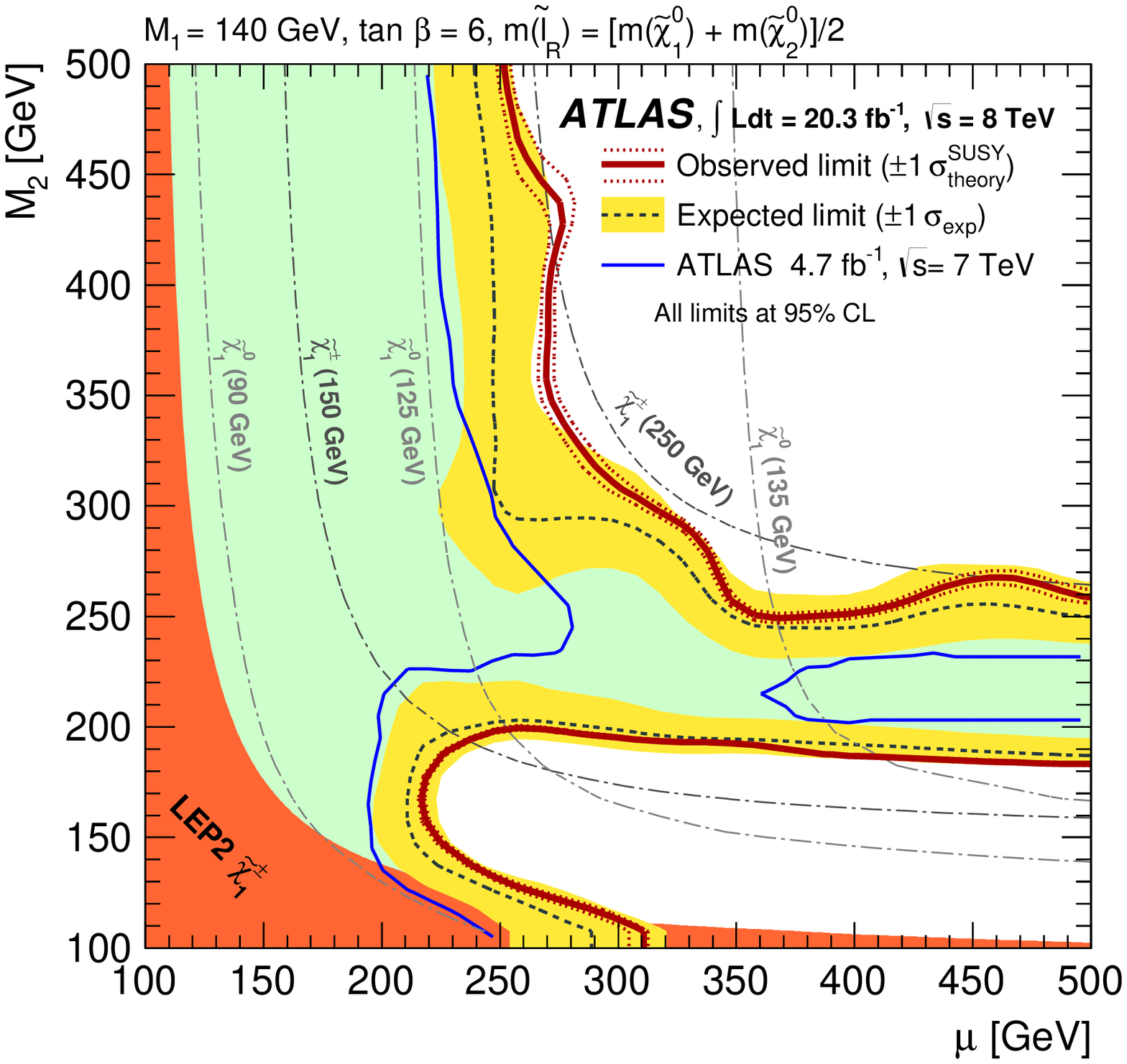}{72,79}
	\raisebox{0.36\textwidth}{(c)} \plot{0.45}{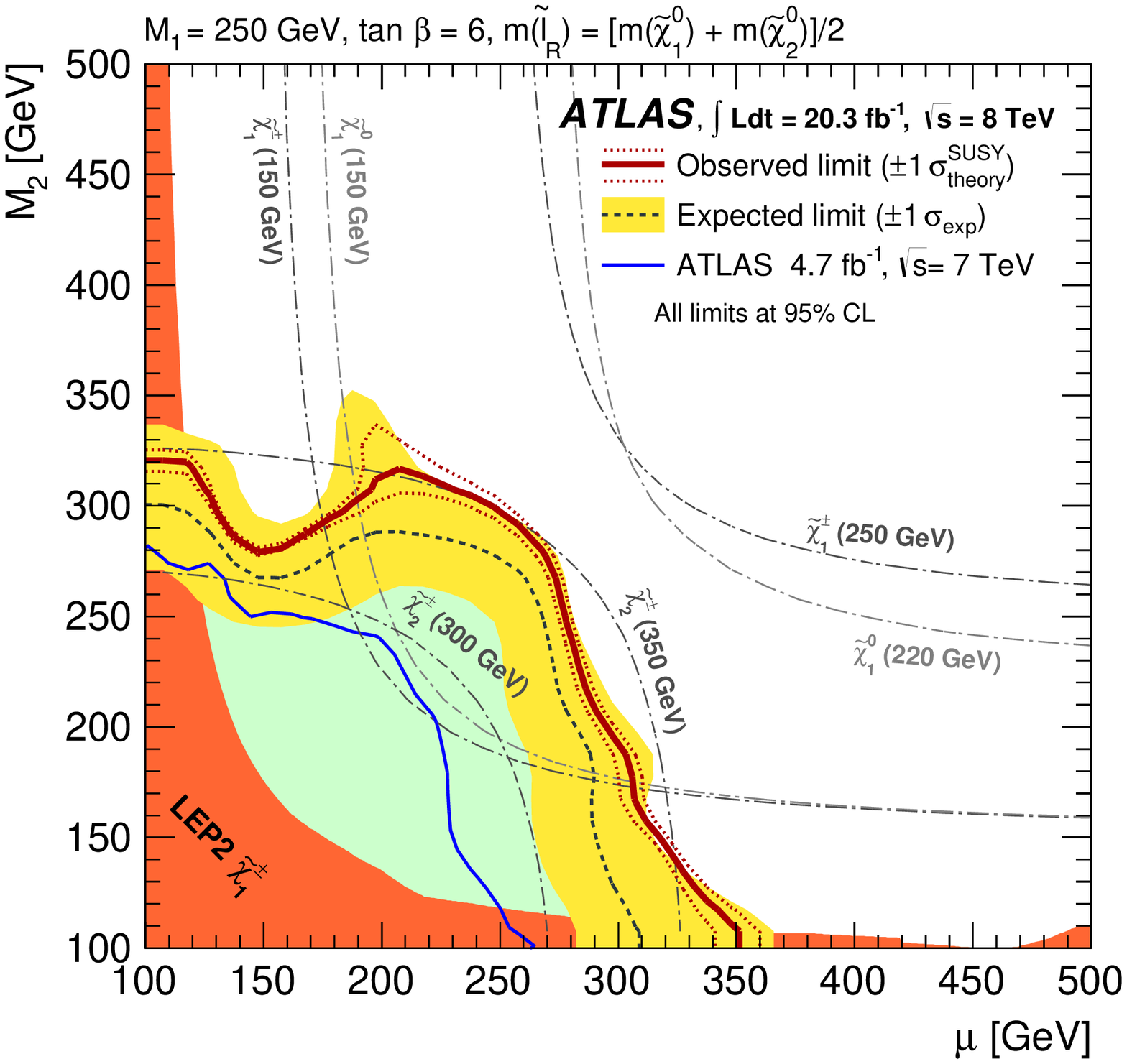}{72,79}
	\raisebox{0.36\textwidth}{(d)} \plot{0.45}{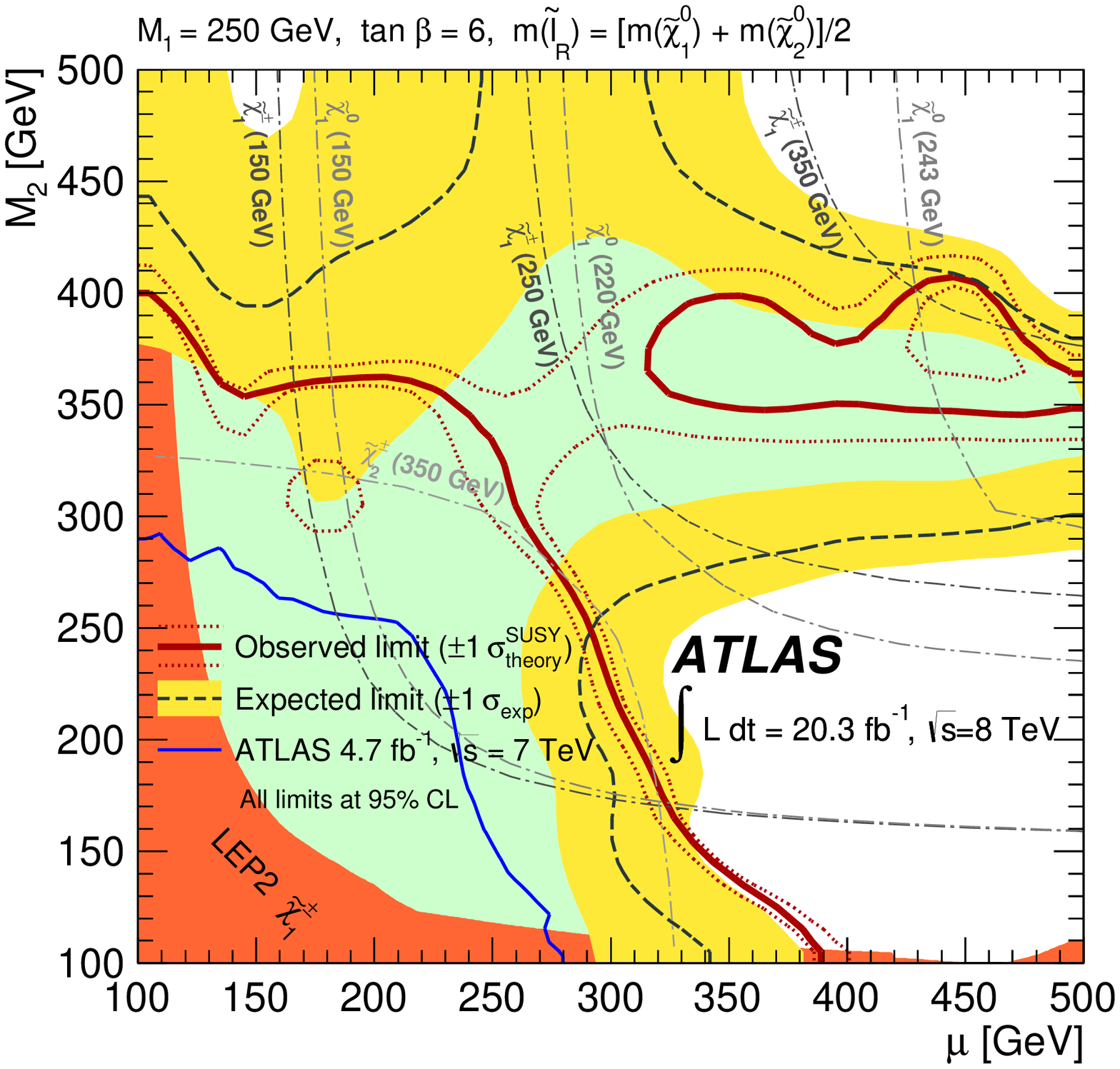}{72,35.5}
\caption{\label{interpretation_pmssm}
	95\% CL exclusion regions in the $\mu$--$M_2$ mass plane of the pMSSM with right-handed slepton
	mass $m_{\smash{\slepton_R}}=(\mNone+\mNtwo)/2$.
	The areas covered by the $-1\sigma$ expected limit are shown in green.
	The $M_1$ parameter is (a) $100\GeV$, (b) $140\GeV$ and (c) $250\GeV$,
	and $\tan\beta=6$.
	The exclusion region for $M_1=250\GeV$ (d) is obtained by combining 
	the results of this analysis
	with those from the ATLAS three-lepton search~\cite{ATLAS_3LEWK2013}.
	The dash-dotted lines indicate the masses of \Cone\ and \None.
	Also shown are the previously reported exclusion regions by ATLAS~\cite{ATLAS_3L7TeV} and the LEP limits~\cite{lepsusy} on the mass of the chargino.
}
\end{figure}

The results are displayed in figures~\ref{interpretation_modeA} through \ref{interpretation_pmssm}.
In each exclusion plot, the solid (dashed) lines show observed (expected) 
exclusion contours, 
including all uncertainties except for the theoretical signal 
cross-section uncertainty arising from the PDF and the renormalization and factorization scales. 
The solid band around the expected exclusion contour shows the $\pm1\sigma$ result where 
all uncertainties, except those on the signal cross-sections, are considered. 
The dotted lines around the observed exclusion contour represent the results obtained 
when varying the nominal signal cross-section by $\pm 1 \sigma$ 
theoretical uncertainty. 
All mass limits hereafter quoted correspond to the 
signal cross-sections reduced by $1\sigma$.

Figure~\ref{interpretation_modeA} shows the 95\% CL exclusion region obtained from \SRmttwo\ 
on the  simplified model for direct \ConeCone\ pair production 
followed by slepton-mediated decays.
For $\mNone=0$,
chargino masses between \exclMconeMinZero\ and \exclMconeMaxZero\ are excluded.
The exclusion in this scenario depends on the assumed slepton mass, 
which is chosen to be halfway between the \chargino{1} and \neutralino{1} masses
in this analysis. 
Studies performed with particle-level signal MC samples show that the signal
acceptance in \SRmttwo\ depends weakly on $\mslepton$, 
and the choice of $\mslepton=(\mCone+\mNone)/2$ minimizes (maximizes) the acceptance
for small (large) \Cone--\None\ mass splitting. 

Figure \ref{fig:WWFinal}(a) shows the 95\% CL exclusion regions obtained from \SRWW\ 
on the simplified-model \ConeCone\ production followed by $W$-mediated decays.
Figure \ref{fig:WWFinal}(b) shows the observed and expected 95\% CL upper limits
on the SUSY signal cross-section normalized by the simplified-model prediction 
as a function of \mCone\ for a massless \neutralino{1}.
For $\mNone=0$, chargino mass ranges of 100--$105\GeV$, 120--$135\GeV$ and 145--$160\GeV$ 
are excluded at 95\% CL.

Figure~\ref{interpretation_modeC}(a) shows the 95\% CL exclusion region obtained from \SRZjets\ 
in the simplified-model \chargino{1}\neutralino{2} production followed by $W$ and $Z$ decays. 
For $\mNone=0$, degenerate \chargino{1} and \neutralino{2} masses 
between \exclMntwoMinZero\ and \exclMntwoMaxZero\ are excluded. 
Figure~\ref{interpretation_modeC}(b) shows the exclusion region obtained by combining
this result with results from the relevant signal regions (SR0a/SR1a/SR1SS/SR2a) 
in the ATLAS search for  electroweak SUSY production in the three-lepton final 
states~\cite{ATLAS_3LEWK2013}. 
The fit is performed on the combined likelihood function using all signal regions.  
The uncertainties are profiled in the likelihood and correlations between channels and processes are taken into account. 
The combination significantly improves the sensitivity. 
As a result, degenerate  \chargino{1} and \neutralino{2} masses between \exclMntwoMinZeroComb\ and \exclMntwoMaxZeroComb\ are excluded at 95\% CL for $\mNone=0$.

Figure~\ref{interpretation_slepton} shows the 95\% CL exclusion regions obtained from \SRmttwo\ for 
the direct production of (a) right-handed, (b) left-handed, and (c) both right- and left-handed
selectrons and smuons of equal mass in the $\mNone$--$m_{\slepton}$ plane.
For $\mNone=0$, 
common values for left and right-handed selectron and 
smuon mass between \exclMslepMinZero\ and \exclMslepMaxZero\ are excluded. 
The sensitivity decreases as the \slepton--\None\ mass splitting decreases
because the \mttwo\ end point of the SUSY signal moves lower towards 
that of the SM background.
For $\mNone=100\GeV$, common left and right-handed 
slepton masses between \exclMslepMinHundred\ and \exclMslepMaxHundred\ are excluded. 
The present result cannot be directly compared with the previous ATLAS slepton limits~\cite{Aad:2012pxa},
which used a flavour-blind signal region and searched for a single slepton flavour with both right-handed and
left-handed contributions.

\begin{figure}\centering
	\raisebox{0.36\textwidth}{(a)} \plot{0.45}{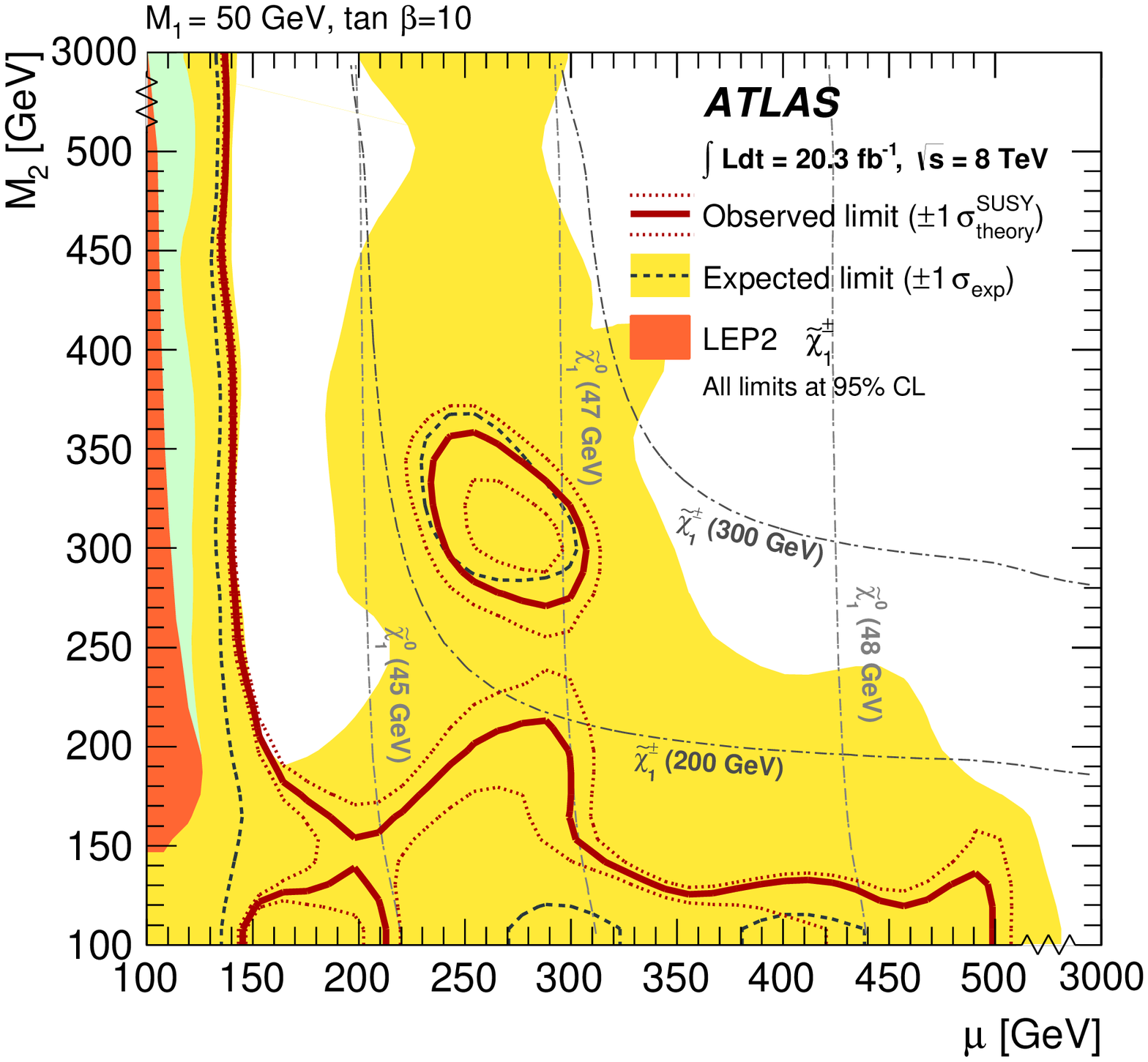}{72,79.5}
	\raisebox{0.36\textwidth}{(b)} \plot{0.45}{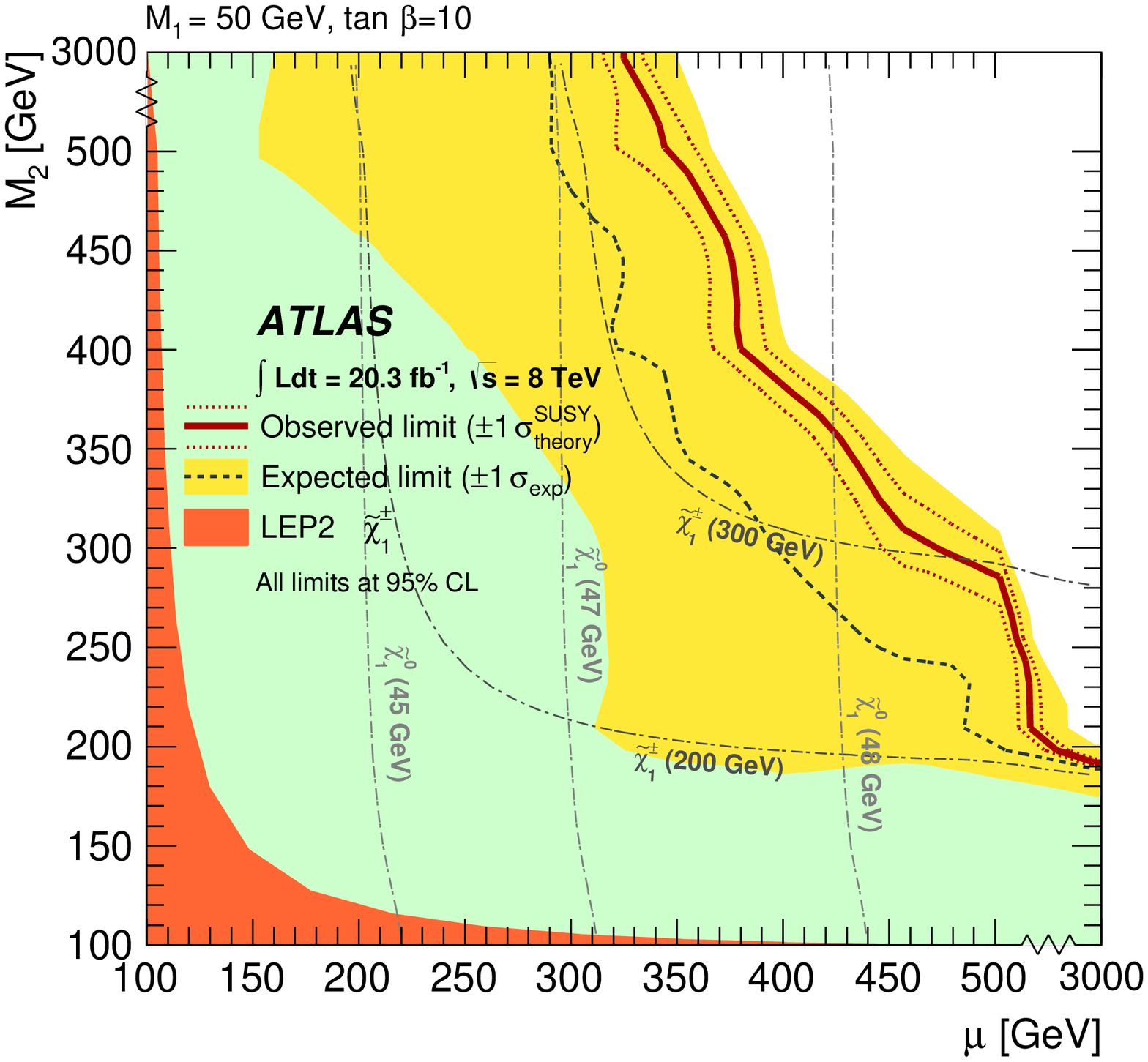}{35,61.5}
\caption{\label{interpretation_pmssm_noslep}
	(a) 95\% CL exclusion regions in the $\mu$--$M_2$ mass plane of the pMSSM with 
	very large slepton masses, $M_1=50\GeV$ and $\tan\beta=10$. 
	(b) The exclusion region obtained by combining the results form \SRZjets\
	with those from the ATLAS three-lepton search~\cite{ATLAS_3LEWK2013}.
	The areas covered by the $-1\sigma$ expected limit are shown in green.
	The dash-dotted lines indicate the masses of \Cone\ and \None.
    Also shown are the LEP limits~\cite{lepsusy} on the mass of the chargino.
}
\end{figure}

Figure~\ref{interpretation_pmssm}(a)--(c) show the 95\% CL exclusion regions in the pMSSM 
$\mu-M_{2}$ plane for the scenario with right-handed sleptons 
with $m_{\smash{\slepton_R}}=(\mNone+\mNtwo)/2$.
The $M_1$ parameter is set to (a) $100\GeV$, (b) $140\GeV$ and (c) $250\GeV$,
and $\tan{\beta}=6$.
At each model point, the limits are obtained using the SR with the best expected sensitivity.
Fig.~\ref{interpretation_pmssm}(d) shows the exclusion region for $M_1=250\GeV$ obtained 
by combining the results of this analysis
with the ATLAS three-lepton results~\cite{ATLAS_3LEWK2013}. 
Figure~\ref{interpretation_pmssm_noslep}(a) shows the 95\% CL exclusion region in the pMSSM 
$\mu-M_{2}$ plane for the scenario with heavy sleptons, $\tan\beta=10$ and $M_1=50\GeV$,
using the SR with the best expected sensitivity at each model point.
The island of exclusion near the centre of figure~\ref{interpretation_pmssm_noslep}(a) 
is due to \SRZjets, and is shaped by the kinematical thresholds of the $\Cone\to W\None$ 
and $\Ntwo\to Z\None$ decays. 
Figure~\ref{interpretation_pmssm_noslep}(b) shows the exclusion region obtained
by combining the results from \SRZjets\ with the three-lepton results. 
These results significantly extend previous limits in the pMSSM $\mu-M_{2}$ plane.

The \CLs\ value is also calculated from \SRWWa\ for the GMSB model point where the chargino is the NLSP
with $\mCone=110\GeV$, $\mNone=113\GeV$ and $\mNtwo=130\GeV$~\cite{Meade:6888v2}. 
The observed and expected \CLs\ values are found to be 0.19 and 0.29, respectively. 
The observed and expected 95\% CL limits on the signal cross-section are 
1.58 and 1.90 times the model prediction, respectively.

\section{Conclusion}

Searches for the electroweak production of charginos, neutralinos and sleptons in final 
states characterized by the presence of two leptons (electrons and muons) and missing 
transverse momentum are performed using \datalumi\ of proton-proton collision data 
at $\sqrt{s}=8\TeV$ recorded with the ATLAS experiment at the Large Hadron Collider. 
No significant excess beyond Standard Model expectations is observed. 
Limits are set on the masses of the lightest chargino \chargino{1}, 
next-to-lightest neutralino \neutralino{2} and 
sleptons for different masses of the lightest neutralino \neutralino{1} in simplified models.
In the scenario of \ConeCone\ pair production with \chargino{1} decaying into \neutralino{1} 
via an intermediate slepton with mass halfway between the \chargino{1} and \neutralino{1}, 
\chargino{1} masses between 
\exclMconeMinZero\ and \exclMconeMaxZero\ are excluded at 95\% CL for a massless \neutralino{1}.  
In the scenario of \ConeCone\ pair production with \chargino{1} decaying into \neutralino{1}
and a $W$ boson, \chargino{1} masses in the ranges
100--$105\GeV$, 120--$135\GeV$ and 145--$160\GeV$ are
excluded at 95\% CL for a massless \neutralino{1}.
This is the first limit for this scenario obtained at a hadron collider.
Finally, in the scenario of \chargino{1}\neutralino{2} production 
with \chargino{1} decaying into $W\neutralino{1}$ and  
\neutralino{2} decaying into $Z\neutralino{1}$, 
common \chargino{1} and \neutralino{2} masses between \exclMntwoMinZero\ and \exclMntwoMaxZero\
are excluded at 95\% CL for a massless \neutralino{1}.
Combining this result with those from ref.~\cite{ATLAS_3LEWK2013} extends the exclusion region to
between \exclMntwoMinZeroComb\ and \exclMntwoMaxZeroComb\@.
In scenarios where sleptons decay directly into \neutralino{1} and a charged lepton, common values 
for left and right-handed slepton masses between \exclMslepMinZero\ and \exclMslepMaxZero\ are excluded 
at 95\% CL for a massless \neutralino{1}. 
Improved exclusion regions are also obtained in the pMSSM $\mu$--$M_2$ plane for four sets of
slepton mass, $M_1$ and $\tan\beta$ values.


\section{Acknowledgements}

We thank CERN for the very successful operation of the LHC, as well as the
support staff from our institutions without whom ATLAS could not be
operated efficiently.

We acknowledge the support of ANPCyT, Argentina; YerPhI, Armenia; ARC,
Australia; BMWF and FWF, Austria; ANAS, Azerbaijan; SSTC, Belarus; CNPq and FAPESP,
Brazil; NSERC, NRC and CFI, Canada; CERN; CONICYT, Chile; CAS, MOST and NSFC,
China; COLCIENCIAS, Colombia; MSMT CR, MPO CR and VSC CR, Czech Republic;
DNRF, DNSRC and Lundbeck Foundation, Denmark; EPLANET, ERC and NSRF, European Union;
IN2P3-CNRS, CEA-DSM/IRFU, France; GNSF, Georgia; BMBF, DFG, HGF, MPG and AvH
Foundation, Germany; GSRT and NSRF, Greece; ISF, MINERVA, GIF, I-CORE and Benoziyo Center,
Israel; INFN, Italy; MEXT and JSPS, Japan; CNRST, Morocco; FOM and NWO,
Netherlands; BRF and RCN, Norway; MNiSW and NCN, Poland; GRICES and FCT, Portugal; MNE/IFA, Romania; MES of Russia and ROSATOM, Russian Federation; JINR; MSTD,
Serbia; MSSR, Slovakia; ARRS and MIZ\v{S}, Slovenia; DST/NRF, South Africa;
MINECO, Spain; SRC and Wallenberg Foundation, Sweden; SER, SNSF and Cantons of
Bern and Geneva, Switzerland; NSC, Taiwan; TAEK, Turkey; STFC, the Royal
Society and Leverhulme Trust, United Kingdom; DOE and NSF, United States of
America.

The crucial computing support from all WLCG partners is acknowledged
gratefully, in particular from CERN and the ATLAS Tier-1 facilities at
TRIUMF (Canada), NDGF (Denmark, Norway, Sweden), CC-IN2P3 (France),
KIT/GridKA (Germany), INFN-CNAF (Italy), NL-T1 (Netherlands), PIC (Spain),
ASGC (Taiwan), RAL (UK) and BNL (USA) and in the Tier-2 facilities
worldwide.

\bibliographystyle{JHEP}
\bibliography{DGSlepton2Lep}

\clearpage
\onecolumn
\input{atlas_authlist} 

\end{document}

%% file: atlas_authlist.tex
\begin{flushleft}
{\Large The ATLAS Collaboration}

\bigskip

G.~Aad$^{\rm 84}$,
B.~Abbott$^{\rm 112}$,
J.~Abdallah$^{\rm 152}$,
S.~Abdel~Khalek$^{\rm 116}$,
O.~Abdinov$^{\rm 11}$,
R.~Aben$^{\rm 106}$,
B.~Abi$^{\rm 113}$,
M.~Abolins$^{\rm 89}$,
O.S.~AbouZeid$^{\rm 159}$,
H.~Abramowicz$^{\rm 154}$
% ...
\bigskip
\\
% ...
$^{11}$ Institute of Physics, Azerbaijan Academy of Sciences, Baku, Azerbaijan\\
% ...
$^{84}$ CPPM, Aix-Marseille Universit{\'e} and CNRS/IN2P3, Marseille, France\\
% ...
$^{89}$ Department of Physics and Astronomy, Michigan State University, East Lansing MI, United States of America\\
% ...
$^{106}$ Nikhef National Institute for Subatomic Physics and University of Amsterdam, Amsterdam, Netherlands\\
% ...
$^{112}$ Homer L. Dodge Department of Physics and Astronomy, University of Oklahoma, Norman OK, United States of America\\
$^{113}$ Department of Physics, Oklahoma State University, Stillwater OK, United States of America\\
% ...
$^{116}$ LAL, Universit{\'e} Paris-Sud and CNRS/IN2P3, Orsay, France\\
% ...
$^{152}$ Institute of Physics, Academia Sinica, Taipei, Taiwan\\
% ...
$^{154}$ Raymond and Beverly Sackler School of Physics and Astronomy, Tel Aviv University, Tel Aviv, Israel\\
% ...
$^{159}$ Department of Physics, University of Toronto, Toronto ON, Canada\\
% ...
\end{flushleft}
